**<u>The difference between Dirac's hole theory and quantum field theory</u>**

by

Dan Solomon

Rauland-Borg Corporation
3450 W. Oakton
Skokie, IL 60076

Email: dan.solomon@rauland.com
Phone: 847-324-8337

May 4, 2003

**<u>Abstract</u>**

Dirac's hole theory and quantum field theory are generally thought to be equivalent.  In fact field theory can be derived from hole theory through the process of second quantization.  However, it can be shown that problems worked in both theories yield different results.  The reason for the difference between the two theories will be examined and the effect that this difference has on the way calculations are done in quantum theory will be discussed.

## **I. Introduction**

It is generally assumed that Dirac hole theory and quantum field theory are identical. However, it has recently been shown by Coutinho et al [1][2] that this is not necessarily the case. Coutinho examined the shift in the vacuum energy due to a time independent perturbing potential. It was shown that hole theory and field theory yield different results when this problem was solved using time independent perturbation theory. In this article we will examine hole theory and field theory in order to understand why these differences occur. To avoid mathematical complications we will consider a simple quantum theory where the electrons are non-interacting and respond to an external classical electric potential.

Hole theory was original proposed by Dirac to deal with the problem of negative energy solutions of the Dirac equation. In hole theory the negative energy states are all occupied by a single electron. The electrons in the negative energy "sea" have all the properties of positive energy electrons in that they respond to an electric field and obey the Dirac equation. If one of these negative energy electrons makes a transition to a positive energy states it leaves a "hole" in the negative energy sea which can be identified with a positron. The conceptual difficulty with hole theory is that it includes an infinite number of negative energy electrons which are not directly observable in their unperturbed state. Only changes from the unperturbed state are observable. In field theory the infinite number of electrons that make up the negative energy sea are replaced by the vacuum state vector $|0\rangle$ where $|0\rangle$ is simply the state for which no electrons or positrons are present.    Thus in field theory the definition of the vacuum is considerably simplified.

In Section II of this discussion we consider the Dirac equation for a single electron in the presence of an external, classical, electric potential.  We will show the solutions to the Dirac equation have certain properties.  They obey the continuity equation and are gauge invariant.  Next, in Section III, we will consider N-electron theory where the electrons are non-interacting.  The N-electron wave function must be anti-symmetric so that the Pauli exclusion principle is obeyed.  It will be shown that the N-electron wave function can be considered to be composed of N individual electrons each of which evolves independently in time according to the single particle Dirac equation.  The total charge, current, or energy is just the sum of the charge, current, or energy, respectively, of each of the individual electrons.  Therefore all the symmetries and properties that are associated with the single particle Dirac equation also apply to N-electron theory.  Therefore N-electron theory is gauge invariant and obeys the continuity equation.

In Section IV and V we show how the Schrödinger representation of quantum field theory can be derived from N-electron theory.  The first part of this process is essentially equivalent to a change in notation.  We go from a cumbersome notation where the N-electron wave function is represented by Slater determinants to a simpler notation using field operators and state vectors.  The second part of this process is to treat the "hole" in the infinite electron vacuum sea as a positron.  We introduce electron and positron creation operators which act on the vacuum state to produce electrons and positrons, respectively.  The concept of a vacuum has, then, been simplified from consisting of an infinite number of negative energy electrons to being the state in which no positive energy electrons or positrons exist. It is shown that the vacuum state is the state for which the free field energy is a minimum. The free field energy is the energy when the electric potential is zero.

The discussion up, to this point, is simply a review of basic quantum mechanics that is generally available in many textbooks. It is assumed that the reader is already somewhat familiar with this material therefore formal proofs are generally not given but references are made to more detailed discussions. The purpose is to present the generally accepted argument that shows that hole theory and the Schrödinger representation of quantum field theory are formally equivalent. The discussion that follows this review contains new and, largely, original material which shows that the two theories are actually not equivalent.

In Sections VI, VII, and VIII we examine the continuity equation and gauge invariance in field theory. Due to the fact that field theory is derived from N-electron theory we expect that gauge invariance and the continuity equation will hold true for field theory since they hold true for N-electron theory. However, it will be shown that there is an inconsistency in field theory. If the continuity equation and gauge invariance are true then there must exist quantum states with less free field energy than the vacuum state. However, if we use the standard definition of the vacuum state, this is not possible. When we examine what is required for the continuity equation and gauge invariance to hold in field theory, it is shown that a quantity called the Schwinger term must be zero. However, as has been shown by Schwinger, this quantity cannot be zero if there are no quantum states with less free field energy than the vacuum state.

Next, in Section IX, we show how this problem with gauge invariance manifests itself in field theory. This problem shows up when effect of an electromagnetic field on the vacuum is considered. An electromagnetic field perturbs the vacuum and produces a vacuum current which can be calculated using perturbation theory. It is well know that when the vacuum current is calculated the result is not gauge invariant. The non-gauge invariant terms that appear in the calculation of the vacuum current must be removed to obtain a physically acceptable result. It is shown that this problem with the gauge invariance of the vacuum current is directly related to the fact that the Schwinger term is non-zero.

In Section X and XI we take on the question on whether the vacuum state in hole theory is the state with the lowest free field energy. This vacuum state is the physical state where all the negative energy states are occupied by a single electron. These electrons obey the Dirac equation and their energy will change in response to an electric field. If an electromagnetic field is applied and then removed, the resulting perturbed vacuum state will have a different energy from the original state. The change in the energy is just the sum of the change in the energies of each of the individual electrons. It is shown in Section X that it is possible to find an electric field which causes the change in the vacuum energy to be negative. That is, energy is extracted from the vacuum due to its interaction with the electric field. This is not possible in field theory because the vacuum is the lowest energy state. This example demonstrates that field theory and hole theory are not equivalent.

The difficulty with this result is that we can no longer assume that the symmetries and properties associated with the single particle Dirac equation are valid for field theory.

In particular, it will be shown that that field theory, as currently formulated, is not gauge invariant and the continuity equation is not true. However, as will be shown in Section XII, the vacuum state can be redefined so that quantum field theory can be made equivalent to hole theory. When this is done the principle of gauge invariance and the continuity equation will be valid in quantum field theory. Throughout this discussion we assume "natural units" so that $\hbar = c = 1$.

## II. The single particle Dirac Equation

In this section we will review some of the properties of the solutions of the Dirac equation. In particular we are interested in local change conservation, or the continuity equation, and gauge invariance. The single particle Dirac equation is [3][4][5],

$$i\frac{\partial \psi(\vec{x},t)}{\partial t} = H(\vec{x},t)\psi(\vec{x},t) \tag{2.1}$$

where $\psi(\vec{x},t)$ is the wave function and $H(\vec{x},t)$ is the Hamiltonian operator which is defined by,

$$H(\vec{x},t) = H_0(\vec{x}) - q\left(\vec{\alpha}\cdot\vec{A}(\vec{x},t) - A_0(\vec{x},t)\right) \tag{2.2}$$

where $\left(A_0(\vec{x},t),\vec{A}(\vec{x},t)\right)$ is the electric potential, 'q' is the charge on the electron, and $H_0(\vec{x})$ is the free field Hamiltonian, which is the Hamiltonian when the electric potential is zero. Throughout this discussion we assume that the electric potential is an unquantized classical quantity. $H_0(\vec{x})$ is given by,

$$H_0(\vec{x}) = -i\vec{\alpha}\cdot\vec{\nabla} + \beta m \tag{2.3}$$

where m is the mass and $\vec{\alpha}$ and $\beta$ are the usual 4x4 matrices (see ref. [3]).

If $\psi_1$ and $\psi_2$ are two wave functions obeying (2.1), then it can be easily shown that the following relationship holds,

$$\frac{\partial\left(\psi_1^\dagger \psi_2\right)}{\partial t} + \vec{\nabla} \cdot \left(\psi_1^\dagger \vec{\alpha} \psi_2\right) = 0 \qquad (2.4)$$

where the ' $\dagger$ ' indicates Hermitian conjugate. Integrate the above over all space to obtain,

$$\frac{\partial}{\partial t}\int\left(\psi_1^\dagger \psi_2\right)d\vec{x} = 0 \qquad (2.5)$$

This means that $\int\left(\psi_1^\dagger \psi_2\right)d\vec{x}$ is constant in time. The main result from this is that if $\psi_1$ and $\psi_2$ are initially orthogonal, i.e. $\int\left(\psi_1^\dagger \psi_2\right)d\vec{x} = 0,$ they will be orthogonal for all time. Also if the wave function $\psi$ is initially normalized (i.e., $\int\left(\psi^\dagger \psi\right)d\vec{x} = 1$) it will be normalized for all time.

Next, define the charge expectation value $\rho_e$ and the current expectation value $\vec{J}_e$ as,

$$\rho_e = q\psi^\dagger \psi \ \text{ and } \ \vec{J}_e = q\psi^\dagger \vec{\alpha} \psi \qquad (2.6)$$

where $\psi$ is normalized. Use this in (2.4) to show that the quantities $\rho_e$ and $\vec{J}_e$ obey the continuity equation,

$$\frac{\partial \rho_e}{\partial t} + \vec{\nabla} \cdot \vec{J}_e = 0 \qquad (2.7)$$

This is the quantum mechanical version of corresponding classical equation for the local conservation of electric charge.

An important property that a physical theory should obey is that of gauge invariance [3]. The electromagnetic field is given in terms of the electric potential by,

$$\vec{E} = -\left(\frac{\partial \vec{A}}{\partial t} + \vec{\nabla} A_0\right) \text{ and } \vec{B} = \vec{\nabla} \times \vec{A} \qquad (2.8)$$

A change in the gauge is a change in the electric potential that does not produce a change in the electromagnetic field. Such a change is given by,

$$\vec{A} \to \vec{A}' = \vec{A} - \vec{\nabla}\chi \text{ and } A_0 \to A_0' = A_0 + \frac{\partial \chi}{\partial t} \qquad (2.9)$$

where $\chi(\vec{x}, t)$ is an arbitrary real valued function. For Dirac field theory to be gauge invariant means that a change in the gauge does not produce a change in any measurable quantities. These include the current and charge expectation values. To show that this is the case substitute (2.9) into (2.2) to obtain,

$$H_g = H + q\left(\vec{\alpha} \cdot \vec{\nabla}\chi + \frac{\partial \chi}{\partial t}\right) \qquad (2.10)$$

where H is the original Hamiltonian and $H_g$ is the Hamiltonian after a gauge transformation. Let $\psi$ be the solution of (2.1) and $\psi_g$ be the solution of the gauge transformed equation,

$$i\frac{\partial \psi_g(\vec{x}, t)}{\partial t} = H_g \psi_g(\vec{x}, t) \qquad (2.11)$$

It can be shown that,

$$\psi_g = e^{-iq\chi}\psi \to \psi_g^\dagger = \psi^\dagger e^{iq\chi} \qquad (2.12)$$

When this result is used in (2.6) it is easy to show that the current and charge expectation values are invariant under a gauge transformation.

Now consider the Dirac equation when the electric potential is zero. This may also by referred to as the free field Dirac equation. In this case (2.1) becomes,

$$i\frac{\partial \psi(\vec{x},t)}{\partial t} = H_0(\vec{x})\psi(\vec{x},t) \tag{2.13}$$

The solutions to this equation are of the form,

$$\varphi_n^{(0)}(\vec{x},t) = \varphi_n^{(0)}(\vec{x})e^{-i\lambda_n E_n t}; \ \ \varphi_n^{(0)}(\vec{x}) = u_n e^{i\vec{p}_n\cdot\vec{x}} \tag{2.14}$$

where $u_n$ is a 4-spinor (see page 30-32 of [3]). The $\varphi_n^{(0)}(\vec{x})$ satisfy,

$$H_0\varphi_n^{(0)}(\vec{x}) = \lambda_n E_n \varphi_n^{(0)}(\vec{x}) \tag{2.15}$$

In the above expression the energy eigenvalue is $\lambda_n E_n$ when $E_n$ is given by

$E_n = \sqrt{\vec{p}_n^2 + m^2}$ , $\vec{p}_n$ is the momentum, $\lambda_n = +1$ for positive energy states, and

$\lambda_n = -1$ for negative energy states. The index 'n' is an integer that stands for the triplet

$\left(\lambda_n, \vec{p}_n, s_n\right)$ where $s = \pm 1/2$ is the spin state.

If the electric potential is zero the triplet $\left(\lambda_n, \vec{p}_n, s_n\right)$ does not change as the state

$\varphi_n^{(0)}(\vec{x},t)$ evolves forward in time. If an electric potential is applied then the initial state

$\varphi_n^{(0)}(\vec{x},t_i)$ will evolve into the final state $\varphi_n(\vec{x},t_f)$. In the case of the final state the

index 'n' can no longer be associated with a given triplet $\left(\lambda_n, \vec{p}_n, s_n\right)$. In this case the index 'n' means that the final state $\varphi_n(\vec{x},t_f)$ evolves from the initial state $\varphi_n^{(0)}(\vec{x},t_i)$.

The $\varphi_n^{(0)}(\vec{x})$ are normalized and they satisfy the relationships,

$$\int \varphi_n^{(0)\dagger}(\vec{x})\varphi_m^{(0)}(\vec{x})d\vec{x} = \delta_{nm} \text{ and } \sum_n \left(\varphi_n^{(0)\dagger}(\vec{x})\right)_a \left(\varphi_n^{(0)}(\vec{y})\right)_b = \delta_{ab}\delta(\vec{x}-\vec{y}) \tag{2.16}$$

where 'a' and 'b' are spinor indices (see page 107 of [6]). Due to the above relationships an arbitrary 4-spinor wave function can be expressed as a Fourier sum in terms of the basis states $\varphi_n^{(0)}(\vec{x})$.

As can be seen from the above discussion there are both positive and negative energy solutions to the Dirac equation. Therefore there is nothing to prevent a single particle from making a transition to a negative energy state if it is perturbed by some external influence. In order to deal with this problem Dirac made the assumption that all the negative energy states were occupied by a single electron and invoked the Pauli exclusion principle which states that no more then one electron can occupy a given energy state. The result of this is that we immediately go from a one electron theory to an N-electron theory where $N \rightarrow \infty$ (see the discussion in Chapt 4 of [3] and page 274-275 of [7]).

### III. N-electron theory

To produce an N-electron theory the N-electron wave function $\Psi^N(\vec{x}_1, \vec{x}_2, ..., \vec{x}_N, t)$ must be defined in such a way that the Pauli principle holds and all N particles are indistinguishable. This is done by defining the wave function so that it is antisymmetric under the exchange of the coordinates. The N particle wave function can be expressed as a Slater determinant [8][9][10],

$$\Psi^N(\vec{x}_1, \vec{x}_2, ..., \vec{x}_N, t) = \frac{1}{\sqrt{N!}} \begin{vmatrix} \psi_1(\vec{x}_1, t) & \psi_1(\vec{x}_2, t) & ... & \psi_1(\vec{x}_N, t) \\ \psi_2(\vec{x}_1, t) & \psi_2(\vec{x}_2, t) & \cdots & \psi_2(\vec{x}_N, t) \\ \vdots & \vdots & \vdots & \vdots \\ \psi_N(\vec{x}_1, t) & \psi_N(\vec{x}_2, t) & ... & \psi_N(\vec{x}_N, t) \end{vmatrix} \quad (3.1)$$

where the $\psi_n \left( n = 1, 2, \ldots, N \right)$ are a set of normalized and orthogonal wave functions. A more compact way of writing this is [10],

$$\Psi^N \left( \vec{x}_1, \vec{x}_2, \ldots, \vec{x}_N, t \right) = \frac{1}{\sqrt{N!}} \sum_P (-1)^p P \left( \psi_1 \left( \vec{x}_1, t \right) \psi_2 \left( \vec{x}_2, t \right) \cdots \psi_N \left( \vec{x}_N, t \right) \right) \qquad (3.2)$$

where P is a permutation operator acting on the space coordinates and p is the number of interchanges in P.

This N-electron wave functions obeys the equation,

$$i \frac{\partial}{\partial t} \Psi^N \left( \vec{x}_1, \vec{x}_2, \ldots, \vec{x}_N, t \right) = H^N \left( \vec{x}_1, \vec{x}_2, \ldots, \vec{x}_N, t \right) \Psi^N \left( \vec{x}_1, \vec{x}_2, \ldots, \vec{x}_N, t \right) \qquad (3.3)$$

where $H^N \left( \vec{x}_1, \vec{x}_2, \ldots, \vec{x}_N, t \right)$ is the N-electron Hamiltonian. For non-interacting electrons, which is the case throughout this discussion, the N-electron Hamiltonian is given by,

$$H^N \left( \vec{x}_1, \vec{x}_2, \ldots, \vec{x}_N, t \right) = \sum_{n=1}^N H \left( \vec{x}_n, t \right) \qquad (3.4)$$

where $H \left( \vec{x}_n, t \right)$ is the single particle Hamiltonian defined by (2.2). The single particle Hamiltonian $H \left( \vec{x}_n, t \right)$ acts on the quantities $\psi_j \left( \vec{x}_n, t \right)$ that contain the position coordinate $\vec{x}_n$. From the above discussion we can show that each of the functions $\psi_j \left( \vec{x}_n, t \right)$ in (3.2) satisfy the single particle Dirac equation (2.1).

The expectation value of a single particle operator $O_{op} \left( \vec{x} \right)$ is defined as,

$$O_e = \int \psi^\dagger \left( \vec{x}_1, t \right) O_{op} \left( \vec{x}_1 \right) \psi \left( \vec{x}_1, t \right) d\vec{x}_1 \qquad (3.5)$$

where $\psi \left( \vec{x}_1, t \right)$ is a normalized single particle wave function. Therefore, from (2.6), we have that the single particle charge and current operators are,

$$\rho_{op} = q \delta \left( \vec{x} - \vec{x}_1 \right) \text{ and } \vec{J}_{op} = q \vec{\alpha} \delta \left( \vec{x} - \vec{x}_1 \right) \qquad (3.6)$$

The N-electron operator is given by,

$$O_{op}^N = \sum_{n=1}^N O_{op}\left(\vec{x}_n\right) \tag{3.7}$$

which is just the sum of one particle operators. The expectation value of a normalized N-electron wave function is,

$$O_e^N = \int \Psi^{N\dagger}\left(\vec{x}_1, \vec{x}_2, ..., \vec{x}_N, t\right) O_{op}^N\left(\vec{x}_1, \vec{x}_2, ..., \vec{x}_N\right) \Psi^N\left(\vec{x}_1, \vec{x}_2, ..., \vec{x}_N, t\right) d\vec{x}_1 d\vec{x}_2 ... d\vec{x}_N$$

$$\tag{3.8}$$

This can be shown to be equal to,

$$O_e^N = \sum_{n=1}^N \int \psi_n^\dagger\left(\vec{x}, t\right) O_{op}\left(\vec{x}\right) \psi_n\left(\vec{x}, t\right) d\vec{x} \tag{3.9}$$

That is, the N electron expectation value is just the sum of the single particle expectation values associated with each individual wave function $\psi_n$. Therefore, the charge and current expectation values for the N electron wave function are,

$$\rho_e^N\left(\vec{x}, t\right) = \sum_{n=1}^N q \psi_n^\dagger\left(\vec{x}, t\right) \psi_n\left(\vec{x}, t\right) \tag{3.10}$$

and

$$\vec{J}_e^N\left(\vec{x}, t\right) = \sum_{n=1}^N q \psi_n^\dagger\left(\vec{x}, t\right) \vec{\alpha} \psi_n\left(\vec{x}, t\right) \tag{3.11}$$

In addition, another quantity that we will be interested in later is the free field energy $\xi_f\left(\Psi^N\right)$ of the N-electron state. This is the energy when the electric potential is zero. Based on the above discussion,

$$\xi_f\left(\Psi^N\right) = \sum_{n=1}^N \int \psi_n^\dagger\left(\vec{x}, t\right) H_0 \psi_n\left(\vec{x}, t\right) d\vec{x} \tag{3.12}$$

The result of the above discussion is that we can view the N-electron system as consisting of N independent orthonormal wave functions $\psi_n(\vec{x}, t)$. Each wave function evolves in time according to the single particle Dirac equation (2.1). The total current and charge expectation value is just the sum of the current and charge expectation value associated with each independent wave function. Similarly, the total free field energy is just the sum of the free field energy of each independent wave function. Therefore, the various symmetries and conservation laws that hold for the one electron wave function hold for the N-electron wave function. In particular, we can easily conclude that the continuity equation and gauge invariance hold for the N-electron wave function.

### IV. Second Quantization

Consider equation (3.2). Each of the terms $\psi_n(\vec{x}, t)$ can be expanded in terms of the basis states $\varphi_j^{(0)}(\vec{x})$, which are defined by equation (2.14). Therefore we can write,

$$\psi_n(\vec{x}, t) = \sum_j c_{nj}(t) \varphi_j^{(0)}(\vec{x}) \tag{4.1}$$

where $c_{nj}(t)$ are the Fourier coefficients. Use this in (3.2) to obtain,

$$\Psi^N = \sum_{j_1, j_2, \ldots, j_N} c_{1, j_1}(t) c_{2, j_2}(t) \ldots c_{N, j_N}(t) \frac{1}{\sqrt{N!}} \sum_P (-1)^p P\left(\varphi_{j_1}^{(0)}(\vec{x}_1) \varphi_{j_2}^{(0)}(\vec{x}_2) \cdots \varphi_{j_N}^{(0)}(\vec{x}_N)\right)$$

$$\tag{4.2}$$

where we write $\Psi^N$ instead of $\Psi^N(\vec{x}_1, \vec{x}_2, \ldots, \vec{x}_N, t)$ to shorten the expression. Note that any of the summation terms in which any of the $j_i$'s are equal is zero. We see, then, that we can expand any arbitrary N electron wave function in terms of N-electron basis states defined by,

$$\Phi^N_{j_1, j_2, \ldots, j_N}(\vec{x}_1, \vec{x}_2, \ldots, \vec{x}_N) = \frac{1}{\sqrt{N!}} \sum_P (-1)^P P\left(\varphi^{(0)}_{j_1}(\vec{x}_1) \varphi^{(0)}_{j_2}(\vec{x}_2) \cdots \varphi^{(0)}_{j_N}(\vec{x}_N)\right) \qquad (4.3)$$

Thus for any N-particle state $\Psi^N$ we can write,

$$\Psi^N(\vec{x}_1, \vec{x}_2, \ldots, \vec{x}_N, t) = \sum_{j_1 > j_2 >, \ldots, > j_N} c_{j_1, j_2, \ldots, j_N}(t) \Phi^N_{j_1, j_2, \ldots, j_N}(\vec{x}_1, \vec{x}_2, \ldots, \vec{x}_N) \qquad (4.4)$$

In the above summation none of the $j_i$'s are equal and each ordered set $(j_1, j_2, \ldots, j_N)$ occurs only once. Next define destruction and creation operators which act on these basis states to the change the number of electrons in each state (see [10]). The destruction operator $\hat{a}_j$ turns the basis state $\Phi^N$ into the basis state $\Phi^{N-1}$ by eliminating the single particle wave function $\varphi^{(0)}_j$ from $\Phi^N$. If $\Phi^N$ does not contain $\varphi^{(0)}_j$ then the action of the operator yields 0. Therefore,

$$\begin{aligned}
&\hat{a}_{j_k} \Phi^N_{j_1, j_2, \ldots, j_k, \ldots, j_N}(\vec{x}_1, \vec{x}_2, \ldots, \vec{x}_N) \\
&= (\pm) \Phi^{N-1}_{j_1, j_2, \ldots, j_{k-1}, j_{k-1}, \ldots, j_N}(\vec{x}_1, \vec{x}_2, \ldots, \vec{x}_{N-1}) \text{ if } \Phi^N \text{ contains } \varphi^{(0)}_{j_k}
\end{aligned} \qquad (4.5)$$

where the sign is positive if the number of functions that precede $\varphi^{(0)}_{j_k}$ in $\Phi^N$ is odd. Also,

$$\hat{a}_{j_k} \Phi^N_{j_1, j_2, \ldots, j_N}(\vec{x}_1, \vec{x}_2, \ldots, \vec{x}_N) = 0 \text{ if } \Phi^N \text{ does not contian } \varphi^{(0)}_{j_k} \qquad (4.6)$$

The creation operator $\hat{a}^\dagger_j$ turns the basis state $\Phi^N$ into the basis state $\Phi^{N+1}$ by adding the wave function $\varphi^{(0)}_j$ to the state $\Phi^N$ that does not already include $\varphi^{(0)}_j$. If $\Phi^N$ includes $\varphi^{(0)}_j$ then the result of the action of $\hat{a}^\dagger_j$ on $\Phi^N$ is zero. Therefore we have,

$$\hat{a}^{\dagger}_{j_k} \Phi^N_{j_1, j_2, \ldots, j_{k-1}, j_{k+1}, \ldots, j_N} \left( \vec{x}_1, \vec{x}_2, \ldots, \vec{x}_N \right)$$

$$= \left( \pm \right) \Phi^{N+1}_{j_1, j_2, \ldots, j_{k-1}, j_k, j_{k+1}, \ldots, j_N} \left( \vec{x}_1, \vec{x}_2, \ldots, \vec{x}_{N+1} \right) \text{ if } \Phi^N \text{ does not contain } \varphi^{(0)}_{j_k} \tag{4.7}$$

where the sign is positive if the number of functions that precede $\varphi^{(0)}_{j_k}$ in $\Phi^{N+1}$ is even.

Also

$$\hat{a}^{\dagger}_{j_k} \Phi^N_{j_1, j_2, \ldots, j_k, \ldots, j_N} \left( \vec{x}_1, \vec{x}_2, \ldots, \vec{x}_N \right) = 0 \text{ if } \Phi^N \text{ contains } \varphi^{(0)}_{j_k} \tag{4.8}$$

It can be shown [10] that the operators $\hat{a}^{\dagger}_j$ and $\hat{a}_j$ satisfy the following anti-commutator relationships,

$$\left\{ \hat{a}_j, \hat{a}_k \right\} = 0; \quad \left\{ \hat{a}^{\dagger}_j, \hat{a}^{\dagger}_k \right\} = 0; \quad \left\{ \hat{a}^{\dagger}_j, \hat{a}_k \right\} = \delta_{jk} \tag{4.9}$$

Next define the empty state $\Phi^0$ which contains no particles. This state is defined by,

$$\hat{a}_j \Phi^0 = 0 \text{ and } \hat{a}^{\dagger}_j \Phi^0 = \Phi^1_j = \varphi^{(0)}_j \left( x_1 \right) \tag{4.10}$$

Any N-electron basis state $\Phi^N$ can be produced by acting on $\Phi^0$ with the creation operators,

$$\Phi^N_{j_1, j_2, \ldots, j_N} = \hat{a}^{\dagger}_{j_1} \hat{a}^{\dagger}_{j_2} \ldots \hat{a}^{\dagger}_{j_N} \Phi^0 \tag{4.11}$$

When this result is used in (4.2) along with (4.3) we see that any arbitrary wave function can be expressed in terms of the creation operators. In this case we use bra-ket notation so that we write $\left| \psi^N \left( t \right) \right\rangle$ instead of $\Psi^N \left( \vec{x}_1, \vec{x}_2, \ldots, \vec{x}_N, t \right)$. Therefore in place of (4.4) we have,

$$\left| \psi^N \left( t \right) \right\rangle = \sum_{j_1 > j_2 >, \ldots, > j_N} c_{j_1, j_2, \ldots, j_N} \left( t \right) \hat{a}^{\dagger}_{j_1} \hat{a}^{\dagger}_{j_2} \ldots \hat{a}^{\dagger}_{j_N} \Phi^0 \tag{4.12}$$

Also, the hermitian conjugate is written as,

$$\left\langle \psi^N(t)\right| = \sum_{j_1 > j_2 >, \ldots, > j_N} \Phi^0 \hat{a}_{j_N} \ldots \hat{a}_{j_2} \hat{a}_{j_1} c^*_{j_1, j_2, \ldots, j_N}(t) \qquad (4.13)$$

Now, recall the previous discussion defining the expectation values of the N electron operators $O^N_{op}(\vec{x}_1, \vec{x}_2, \ldots, \vec{x}_N, t)$ that act on the quantities $\psi^N(\vec{x}_1, \vec{x}_2, \ldots, \vec{x}_N, t)$. We want to define operators $\hat{O}$ that act on the quantities $\left|\psi^N(t)\right\rangle$ and correspond to the operator $O^N_{op}(\vec{x}_1, \vec{x}_2, \ldots, \vec{x}_N, t)$ in that the expectation values are the same in both cases. The expectation value associated with some operator $\hat{O}$ for $\left|\psi^N(t)\right\rangle$ is defined by,

$$O_e = \left\langle \psi^N \left| \hat{O} \right| \psi^N \right\rangle \qquad (4.14)$$

In order to that this be the same as the expectation for $\psi^N(\vec{x}_1, \vec{x}_2, \ldots, \vec{x}_N, t)$ given by equation (3.9), the operator $\hat{O}$, that corresponds to $O^N_{op}(\vec{x}_1, \vec{x}_2, \ldots, \vec{x}_N, t)$, must be defined by,

$$\hat{O} = \sum_{i,j} \hat{a}^\dagger_i \hat{a}_j \int \varphi^{(0)\dagger}_i(\vec{x}) O_{op}(\vec{x}) \varphi^{(0)}_j(\vec{x}) d\vec{x} \qquad (4.15)$$

This can be rewritten as,

$$\hat{O} = \int \hat{\psi}^\dagger(\vec{x}) O_{op}(\vec{x}) \hat{\psi}(\vec{x}) d\vec{x} \qquad (4.16)$$

where $\hat{\psi}(\vec{x})$ is called the field operator and is defined by,

$$\hat{\psi}(\vec{x}) = \sum_j \hat{a}_j \varphi^{(0)}_j(\vec{x}) \quad \text{and} \quad \hat{\psi}^\dagger(\vec{x}) = \sum_j \hat{a}^\dagger_j \varphi^{(0)\dagger}_j(\vec{x}) \qquad (4.17)$$

The current and charge operators then become,

$$\hat{\vec{J}}(\vec{x}) = q \hat{\psi}^\dagger(\vec{x}) \vec{\alpha} \hat{\psi}(\vec{x}) \quad \text{and} \quad \hat{\rho}(\vec{x}) = q \hat{\psi}^\dagger(\vec{x}) \hat{\psi}(\vec{x}) \qquad (4.18)$$

The Hamiltonian operator is given by,

$$\hat{H} = \int \hat{\psi}^\dagger(\vec{x}) H(\vec{x}, t) \hat{\psi}(\vec{x}) d\vec{x} \tag{4.19}$$

The Dirac equation (3.3) becomes,

$$i \frac{\partial |\Omega(t)\rangle}{\partial t} = \hat{H} |\Omega(t)\rangle \text{ and } -i \frac{\partial \langle \Omega(t)|}{\partial t} = \langle \Omega(t)| \hat{H} \tag{4.20}$$

where $|\Omega(t)\rangle$, instead of $|\psi^N\rangle$, is used to designate the state vector.

Use (2.2) in (4.19) to obtain,

$$\hat{H} = \int \hat{\psi}^\dagger(\vec{x}) \Big( H_0(\vec{x}) - q\big(\vec{\alpha} \cdot \vec{A}(\vec{x}, t) - A_0(\vec{x}, t)\big) \Big) \hat{\psi}(\vec{x}) d\vec{x} \tag{4.21}$$

This can be written as,

$$\hat{H} = \hat{H}_0 - \int \hat{J}(\vec{x}) \cdot \vec{A}(\vec{x}, t) d\vec{x} + \int \rho(\vec{x}) \cdot A_0(\vec{x}, t) d\vec{x} \tag{4.22}$$

where we have used (4.18) and define,

$$\hat{H}_0 = \int \hat{\psi}^\dagger(\vec{x}) H_0(\vec{x}) \hat{\psi}(\vec{x}) d\vec{x} \tag{4.23}$$

Use (4.17) along with (2.15) and (2.16) in the above to obtain,

$$\hat{H}_0 = \sum_n \lambda_n E_n \hat{a}_n^\dagger \hat{a}_n \tag{4.24}$$

The above relationships ((4.16) through (4.24)) represent quantum field theory in the Schrödinger representation. The basic elements of the theory are the state vectors $|\Omega(t)\rangle$ and the field operator $\hat{\psi}(\vec{x})$. The state vectors $|\Omega(t)\rangle$ evolve in time according to equation (4.20) while the field operators are constant in time. Operators associated with observables are defined in terms of the field operators (e.g. equation (4.18)). According to the derivation given here quantum field theory is equivalent to N-electron theory. Essentially the cumbersome notation of N-electron theory is replaced by the more compact notion of field theory.

Note that there is an alternative version of field theory that is commonly used called the Heisenberg representation. In this case the state vector is constant in time and the time dependence of the theory is associated with the field operators. It can be shown that both representations have the same expectation values and are, therefore, equivalent.

### V. The vacuum state

The final step in the formulation of quantum field theory is the definition of the vacuum state $|0\rangle$. We start with the hole theory concept that the vacuum state is the state where all negative energy states are occupied and the positive energy states are empty. Order the index 'n' associated with the basis wave function $\varphi_n^{(0)}(\vec{x})$ so that $n > 0$ implies a positive energy state and $n < 0$ implies a negative energy state. Further order the states so that the magnitude of 'n' increases with the magnitude of the energy. Following Greiner[3] the vacuum state in quantum field theory is defined by,

$$|0\rangle = \left(\hat{a}_{-1}^\dagger \hat{a}_{-2}^\dagger \hat{a}_{-3}^\dagger \cdots\right)\Phi^0 = \prod_{n=1}^\infty \hat{a}_{-n}^\dagger \Phi^0 \tag{5.1}$$

Thus all negative energy states are occupied and all positive energy states are unoccupied. From (4.24) it is seen that $|0\rangle$ is an eigenstate of the free field Hamiltonian operator $\hat{H}_0$,

$$\hat{H}_0 |0\rangle = \xi_{vac} |0\rangle \tag{5.2}$$

where the energy eigenvalue $\xi_{vac}$ is given by,

$$\xi_{vac} = \sum_{n=1}^\infty \lambda_{-n} E_{-n} = -\sum_{n=1}^\infty E_{-n} \tag{5.3}$$

Note that $\xi_{vac}$ is an infinite negative number. This does not concern us, at this point, because we are interested in changes of energy from some initial state and not the actual value. Now additional eigenstates $|k\rangle$ can be formed by acting on the vacuum state with the creation operators $\hat{a}_j^\dagger$ or the destruction operators $\hat{a}_{-j}$ where 'j' is a positive nonzero integer. The effect of operating with $\hat{a}_j^\dagger$ is to create an electron with positive energy $\lambda_j E_j$. The effect of operating with $\hat{a}_{-j}$ is to destroy an electron with negative energy $\lambda_{-j} E_{-j}$. In either case the energy of the state is increased.

The final step in the process of going from hole theory to field theory is to replace the operator $\hat{a}_{-j}$, that destroys a negative energy electron, by the operator $\hat{d}_j^\dagger$, that creates a positive energy positron. Thus we make the following change in notation,

$$\hat{b}_j = \hat{a}_j \rightarrow \hat{b}_j^\dagger = \hat{a}_j^\dagger \text{ and } \hat{d}_j = \hat{a}_{-j}^\dagger \rightarrow \hat{d}_j^\dagger = \hat{a}_{-j} \text{ where } j > 0 \qquad (5.4)$$

In the above expressions $\hat{d}_j$ and $\hat{d}_j^\dagger$ are positron destruction and creation operators, respectively. The $\hat{b}_j$ and $\hat{b}_j^\dagger$ are the electron destruction and creation operators, respectively, that act on electrons in positive energy states. Use this notation in (4.17) to obtain,

$$\hat{\psi}(\vec{x}) = \sum_{j=1}^{\infty} \left( \hat{b}_j \varphi_j(\vec{x}) + \hat{d}_j^\dagger \varphi_{-j}(\vec{x}) \right) \qquad (5.5)$$

and,

$$\hat{\psi}^\dagger(\vec{x}) = \sum_{j=1}^{\infty} \left( \hat{b}_j^\dagger \varphi_j^\dagger(\vec{x}) + \hat{d}_j \varphi_{-j}^\dagger(\vec{x}) \right) \qquad (5.6)$$

The free field Hamiltonian operator becomes,

$$\hat{H}_0 = \sum_{n=1}^{\infty} \left( \lambda_n E_n \hat{b}_n^\dagger \hat{b}_n + \lambda_{-n} E_{-n} \hat{d}_n \hat{d}_n^\dagger \right) = \sum_{n=1}^{\infty} \left( E_n \hat{b}_n^\dagger \hat{b}_n - E_n \hat{d}_n \hat{d}_n^\dagger \right) \tag{5.7}$$

where we have used $E_n = E_{-n}$, $\lambda_n = 1$, and $\lambda_{-n} = -1$ for $n > 0$. The destruction and creation operators satisfy,

$$\left\{ \hat{d}_j, \hat{d}_k^\dagger \right\} = \delta_{jk} \; ; \; \left\{ \hat{b}_j, \hat{b}_k^\dagger \right\} = \delta_{jk} \; ; \; \text{all other anti-commutators are zero} \tag{5.8}$$

The vacuum state $|0\rangle$ is now defined by,

$$\hat{d}_j |0\rangle = \hat{b}_j |0\rangle = 0 \; \text{ and } \; \langle 0| \hat{d}_j^\dagger = \langle 0| \hat{b}_j^\dagger = 0 \tag{5.9}$$

Use (5.8) in (5.7) to obtain,

$$\hat{H}_0 = \sum_{n=1}^{\infty} E_n \left( \hat{b}_n^\dagger \hat{b}_n - \left( 1 - \hat{d}_n^\dagger \hat{d}_n \right) \right) = \sum_{n=1}^{\infty} \left( E_n \hat{b}_n^\dagger \hat{b}_n + E_n \hat{d}_n^\dagger \hat{d}_n \right) - \xi_{vac} \tag{5.10}$$

At this point redefine $\hat{H}_0$ by adding the constant $\xi_{vac}$ to obtain,

$$\hat{H}_0 = \sum_{n=1}^{\infty} \left( E_n \hat{b}_n^\dagger \hat{b}_n + E_n \hat{d}_n^\dagger \hat{d}_n \right) \tag{5.11}$$

This last step does not affect any results and simply corresponds to a shift in energy,

making the energy of the vacuum state equal to zero. This will simplify some of the mathematical analysis but does not change any of the results of the following discussion.

We can now think of the vacuum state $|0\rangle$ as the state which contains no electrons, no positrons, and has zero energy. New eigenstates states $|k\rangle$ are created by acting on $|0\rangle$ with electron and positron creation operators, e.g,

$$|k\rangle = \hat{b}_{j_1}^\dagger \hat{b}_{j_2}^\dagger ... \hat{b}_{j_s}^\dagger \hat{d}_{v_1}^\dagger \hat{d}_{v_2}^\dagger ... \hat{d}_{v_r}^\dagger |0\rangle \tag{5.12}$$

From this definition and (5.11) we have,

$$\hat{H}_0 \left| k \right\rangle = \xi_{\left| k \right\rangle} \left| k \right\rangle \tag{5.13}$$

where $\xi_{\left| k \right\rangle}$ is the energy eigenvalue of the eigenstate $\left| k \right\rangle$ and is given by,

$$\xi_{\left| k \right\rangle} = \left( E_{j_1} + E_{j_2} + ... + E_{j_s} \right) + \left( E_{v_1} + E_{v_2} + ... + E_{v_r} \right) \tag{5.14}$$

Since all the quantities $E_j$ and $E_v$ in the above equation are greater than zero we have that,

$$\xi_{\left| k \right\rangle} > \xi_{\left| 0 \right\rangle} = 0 \text{ for all } \left| k \right\rangle \neq \left| 0 \right\rangle \tag{5.15}$$

The eigenstates $\left| k \right\rangle$ form an orthonormal set in fock space and satisfy [11],

$$\left\langle k \middle| q \right\rangle = \delta_{kq} \text{ and } \sum_{\left| k \right\rangle} \left| k \right\rangle \left\langle k \right| = 1 \tag{5.16}$$

where the summation is over all eigenstates $\left| k \right\rangle$. Any arbitrary state vector $\left| \Omega \right\rangle$ can be expressed as an expansion in terms of the eigenstates $\left| k \right\rangle$ as,

$$\left| \Omega \right\rangle = \sum_{\left| k \right\rangle} c_{\left| k \right\rangle} \left| k \right\rangle \tag{5.17}$$

The free field energy of a given normalized state $\left| \Omega \right\rangle$ is defined by,

$$\xi_f \left( \left| \Omega \right\rangle \right) = \left\langle \Omega \middle| \hat{H}_0 \middle| \Omega \right\rangle \tag{5.18}$$

From the above discussion we have that,

$$\xi_f \left( \left| \Omega \right\rangle \right) = \left\langle \Omega \middle| \hat{H}_0 \middle| \Omega \right\rangle = \sum_{\left| k \right\rangle} \left| c_{\left| k \right\rangle} \right|^2 \xi_{\left| k \right\rangle} \tag{5.19}$$

Since the $\xi_{\left| k \right\rangle}$ are all positive we have that,

$$\xi_f \left( \left| \Omega \right\rangle \right) \geq \xi_f \left( \left| 0 \right\rangle \right) = 0 \text{ for all } \left| \Omega \right\rangle \tag{5.20}$$

Therefore the vacuum state $\left| 0 \right\rangle$ represents a lower bound to the free field energy.

# VI. Gauge Invariance and Field Theory

The discussion up to this point has essentially been a review of various elements of quantum theory. These elements, in one form or another, are found in many basic works on quantum mechanics. The purpose of this review was to trace the path from the single particle Dirac equation to N-electron theory and then to the Schrödinger representation of quantum field theory in which it shown that N-electron theory and field theory are equivalent. Hole theory can be considered as N-electron theory with $N \rightarrow \infty$ and the vacuum defined as the state in which all negative energy states are occupied. N-electron theory can be shown to obey the continuity equation and to be gauge invariant. Therefore we expect that quantum field theory should also obey the continuity equation and be gauge invariant. In the remainder of this section we will show that this is not, necessarily, the case (see also ref. [12]).

Consider a normalized state vector $\left| \Omega(t) \right\rangle$. Assume that at some initial time, say $t = t_i$, $\left| \Omega(t_i) \right\rangle$ satisfies (5.20), i.e., $\xi_f \left( \left| \Omega(t_i) \right\rangle \right) \geq \xi_f \left( \left| 0 \right\rangle \right)$. Now let $\left| \Omega(t) \right\rangle$ evolve according to equation (4.20). If the electric potential is non-zero then, in general, the free field energy $\xi_f \left( \left| \Omega(t) \right\rangle \right)$ will change in time. It will be shown that if we assume that field theory is gauge invariant and the continuity equation holds then it is possible to specify an electric potential so that (5.20) is not true at some final time $t_f > t_i$. Therefore if (5.20) is true then the principle of gauge invariance or the continuity equation or both is not valid for field theory.

Start by taking the time derivative of $\xi_f \left( \left| \Omega(t) \right\rangle \right)$ and use (4.20) and (5.18) to obtain,

$$\frac{\partial \xi_f \left( \left| \Omega(t) \right\rangle \right)}{\partial t} = i \left\langle \Omega(t) \right| \left[ \hat{H}, \hat{H}_0 \right] \left| \Omega(t) \right\rangle \tag{6.1}$$

Next use (4.22) and substitute for $\hat{H}_0$ in the above to obtain,

$$\frac{\partial \xi_f \left( \left| \Omega(t) \right\rangle \right)}{\partial t} = i \left\langle \Omega(t) \right| \left[ \hat{H}, \left( \hat{H} + \int \hat{\vec{J}}(\vec{x}) \cdot \vec{A}(\vec{x},t) d\vec{x} - \int \hat{\rho}(\vec{x}) A_0(\vec{x},t) d\vec{x} \right) \right] \left| \Omega(t) \right\rangle \tag{6.2}$$

Rearrange terms to yield,

$$\frac{\partial \xi_f \left( \left| \Omega(t) \right\rangle \right)}{\partial t} = i \begin{pmatrix} \int \left\langle \Omega(t) \right| \left[ \hat{H}, \hat{\vec{J}}(\vec{x}) \right] \left| \Omega(t) \right\rangle \cdot \vec{A}(\vec{x},t) d\vec{x} \\ - \int \left\langle \Omega(t) \right| \left[ \hat{H}, \hat{\rho}(\vec{x}) \right] \left| \Omega(t) \right\rangle A_0(\vec{x},t) d\vec{x} \end{pmatrix} \tag{6.3}$$

The current and charge expectation values are defined by,

$$\vec{J}_e(\vec{x},t) = \left\langle \Omega(t) \right| \hat{\vec{J}}(\vec{x}) \left| \Omega(t) \right\rangle \text{ and } \rho_e(\vec{x},t) = \left\langle \Omega(t) \right| \hat{\rho}(\vec{x}) \left| \Omega(t) \right\rangle \tag{6.4}$$

Use this along with (4.20) to obtain,

$$\frac{\partial \vec{J}_e(\vec{x},t)}{\partial t} = i \left\langle \Omega(t) \right| \left[ \hat{H}, \hat{\vec{J}}(\vec{x}) \right] \left| \Omega(t) \right\rangle \text{ and } \frac{\partial \rho_e(\vec{x},t)}{\partial t} = i \left\langle \Omega(t) \right| \left[ \hat{H}, \hat{\rho}(\vec{x}) \right] \left| \Omega(t) \right\rangle \tag{6.5}$$

Use this result in (6.3) to obtain,

$$\frac{\partial \xi_f \left( \left| \Omega(t) \right\rangle \right)}{\partial t} = \left( \int \frac{\partial \vec{J}_e(\vec{x},t)}{\partial t} \cdot \vec{A}(\vec{x},t) d\vec{x} - \int \frac{\partial \rho_e(\vec{x},t)}{\partial t} A_0(\vec{x},t) d\vec{x} \right) \tag{6.6}$$

Next, since we assume that the continuity equation is true for quantum field theory, refer to (2.7) and substitute for $\partial \rho_e / \partial t$ to obtain,

$$\frac{\partial \xi_f \left( \left| \Omega(t) \right\rangle \right)}{\partial t} = \left( \int \frac{\partial \vec{J}_e(\vec{x},t)}{\partial t} \cdot \vec{A}(\vec{x},t) d\vec{x} + \int \vec{\nabla} \cdot \vec{J}_e(\vec{x},t) A_0(\vec{x},t) d\vec{x} \right) \tag{6.7}$$

Now let the electric potential be given by,

$$\left( A_0\left(\vec{x},t\right), \vec{A}\left(\vec{x},t\right)\right) = \left( \frac{\partial \chi\left(\vec{x},t\right)}{\partial t}, -\vec{\nabla}\chi\left(\vec{x},t\right)\right) \qquad (6.8)$$

where $\chi\left(\vec{x},t\right)$ is an arbitrary real valued function. Use this in (6.7) to obtain,

$$\frac{\partial \xi_f\left(\left|\Omega(t)\right\rangle\right)}{\partial t} = \left( -\int \frac{\partial \vec{J}_e\left(\vec{x},t\right)}{\partial t}\cdot \vec{\nabla}\chi\left(\vec{x},t\right)d\vec{x} + \int \vec{\nabla}\cdot\vec{J}_e\left(\vec{x},t\right)\frac{\partial \chi\left(\vec{x},t\right)}{\partial t}d\vec{x}\right) \qquad (6.9)$$

Integrate the first integral by parts and assume reasonable boundary conditions to obtain,

$$\frac{\partial \xi_f\left(\left|\Omega(t)\right\rangle\right)}{\partial t} = \left( \int \frac{\partial \vec{\nabla}\cdot\vec{J}_e\left(\vec{x},t\right)}{\partial t}\chi\left(\vec{x},t\right)d\vec{x} + \int \vec{\nabla}\cdot\vec{J}_e\left(\vec{x},t\right)\frac{\partial \chi\left(\vec{x},t\right)}{\partial t}d\vec{x}\right) \qquad (6.10)$$

This becomes,

$$\frac{\partial \xi_f\left(\left|\Omega(t)\right\rangle\right)}{\partial t} = \frac{\partial}{\partial t}\int \chi\left(\vec{x},t\right)\vec{\nabla}\cdot\vec{J}_e\left(\vec{x},t\right)d\vec{x} \qquad (6.11)$$

Integrate the above from the initial time $t = t_i$ to some final time $t_f > t_i$. This yields,

$$\xi_f\left(\left|\Omega(t_f)\right\rangle\right) - \xi_f\left(\left|\Omega(t_i)\right\rangle\right) = \int \chi\left(\vec{x},t_f\right)\vec{\nabla}\cdot\vec{J}_e\left(\vec{x},t_f\right)d\vec{x} - \int \chi\left(\vec{x},t_i\right)\vec{\nabla}\cdot\vec{J}_e\left(\vec{x},t_i\right)d\vec{x} \qquad (6.12)$$

Next invoke the principle of gauge invariance. When the electric potential of (6.8) is substituted into (2.8) it can be seen that the electromagnetic field is zero for all functions $\chi\left(\vec{x},t\right)$. Therefore, from the assumption of gauge invariance, the current expectation value $\vec{J}_e\left(\vec{x},t\right)$ is independent of $\chi\left(\vec{x},t\right)$. This means that $\chi\left(\vec{x},t\right)$ can be varied in an arbitrary manner without affecting $\vec{J}_e\left(\vec{x},t\right)$. Therefore, if $\vec{\nabla}\cdot\vec{J}_e\left(\vec{x},t_f\right)$ is non-zero then it is always possible to find a $\chi\left(\vec{x},t\right)$ which makes the final free field energy, $\xi_f\left(\left|\Omega(t_f)\right\rangle\right)$, a negative number with an arbitrarily large magnitude. For example, let $\chi\left(\vec{x}, t_i\right) = 0$ and $\chi\left(\vec{x}, t_f\right) = -f\vec{\nabla}\cdot\vec{J}_e\left(\vec{x}, t_f\right)$ where f is a constant. Use this in (6.12) to obtain,

$$\xi_f\left(\left|\Omega\left(t_f\right)\right\rangle\right) = \xi_f\left(\left|\Omega\left(t_i\right)\right\rangle\right) - f\int\left(\vec{\nabla}\cdot\vec{J}_e\left(\vec{x}, t_f\right)\right)^2 d\vec{x} \qquad (6.13)$$

If $\vec{\nabla}\cdot\vec{J}_e\left(\vec{x}, t_f\right)$ is non-zero, the integral on the right is always positive so that as $f\to\infty$, $\xi_f\left(\left|\Omega\left(t_f\right)\right\rangle\right)\to-\infty$. Therefore there is no lower bound for the free field energy and the relationship given in (5.20) cannot be true for all state vectors $\left|\Omega\left(t\right)\right\rangle$.

Now the above result depends on the assumption that the divergence of the current expectation value is non-zero. How do we know that this will be the case? If we assume that quantum theory is a correct model of the real world then we can always find a quantum state for which $\vec{\nabla}\cdot\vec{J}_e\left(\vec{x}, t_f\right)$ is non-zero because there are many examples in the real world where this is, indeed, the case.

In the discussion leading up to (6.13) we have used equation (4.20) for the evolution of the state vector $\left|\Omega\left(t\right)\right\rangle$, assumed that the continuity equation is true, and assumed that the theory is gauge invariant. We find that these conditions are not consistent with the relationship given by (5.20). Therefore, one or more of these relationships must not be valid.

## VII. The Schwinger Term

In the previous section we assumed that the continuity equation was true. The basis of this assumption was that quantum field theory is derived from N-electron theory and that the continuity equation is certainly true for N-electron theory. In this section will we will discuss the requirements for continuity equation to be true in the Schrödinger representation of quantum field theory. The expression for the continuity equation is given by (2.7). Use this and (6.4) to obtain,

$$\frac{\partial \langle \Omega(t) | \hat{\rho}(\vec{x}) | \Omega(t) \rangle}{\partial t} = -\langle \Omega(t) | \vec{\nabla} \cdot \hat{\vec{J}}(\vec{x}) | \Omega(t) \rangle \tag{7.1}$$

Next use (4.20) in the above to obtain,

$$i \langle \Omega(t) | \left[ \hat{H}, \hat{\rho}(\vec{x}) \right] | \Omega(t) \rangle = -\langle \Omega(t) | \vec{\nabla} \cdot \hat{\vec{J}}(\vec{x}) | \Omega(t) \rangle \tag{7.2}$$

Use (4.22) in the above to yield,

$$i \langle \Omega(t) | \left( \begin{array}{c} \left[ \hat{H}_0, \hat{\rho}(\vec{x}) \right] - \int \left[ \hat{\vec{J}}(\vec{y}), \hat{\rho}(\vec{x}) \right] \cdot \vec{A}(\vec{y}, t) d\vec{y} \\ + \int \left[ \hat{\rho}(\vec{y}), \hat{\rho}(\vec{x}) \right] A_0(\vec{y}, t) d\vec{y} \end{array} \right) | \Omega(t) \rangle = -\langle \Omega(t) | \vec{\nabla} \cdot \hat{\vec{J}}(\vec{x}) | \Omega(t) \rangle \tag{7.3}$$

For the above equation to be true for arbitrary values of the state vector $| \Omega(t) \rangle$ and the electric potential $\left( A_0(\vec{x}, t), \vec{A}(\vec{x}, t) \right)$ the following relationships must hold,

$$\left[ \hat{\vec{J}}(\vec{y}), \hat{\rho}(\vec{x}) \right] = 0 \tag{7.4}$$

$$i \left[ \hat{H}_0, \hat{\rho}(\vec{x}) \right] = -\vec{\nabla} \cdot \hat{\vec{J}}(\vec{x}) \tag{7.5}$$

$$\left[ \hat{\rho}(\vec{y}), \hat{\rho}(\vec{x}) \right] = 0 \tag{7.6}$$

These relationships are a sufficient and necessary condition for the continuity equation to be true in Dirac field theory. However it has been show by Schwinger [13] that if (5.15) and (7.5) are true then (7.4) cannot be true. Define the Schwinger term by

$$ST(\vec{y}, \vec{x}) = \left[ \hat{\rho}(\vec{y}), \hat{\vec{J}}(\vec{x}) \right] \tag{7.7}$$

According to (7.4) $ST(\vec{y}, \vec{x})$ must be zero for the continuity equation to be valid.

Take the divergence of the Schwinger term $\left[\hat{\rho}(\vec{y}),\hat{J}(\vec{x})\right]$ and use (7.5) to obtain,

$$\vec{\nabla}_{\vec{x}}\cdot\left[\hat{\rho}(\vec{y}),\hat{\vec{J}}(\vec{x})\right]=\left[\hat{\rho}(\vec{y}),\vec{\nabla}\cdot\hat{\vec{J}}(\vec{x})\right]=-i\left[\hat{\rho}(\vec{y}),\left[\hat{H}_0,\hat{\rho}(\vec{x})\right]\right] \tag{7.8}$$

Next expand the commutator to yield,

$$i\vec{\nabla}_{\vec{x}}\cdot\left[\hat{\rho}(\vec{y}),\hat{\vec{J}}(\vec{x})\right]=-\hat{H}_0\hat{\rho}(\vec{x})\hat{\rho}(\vec{y})+\hat{\rho}(\vec{x})\hat{H}_0\hat{\rho}(\vec{y})+\hat{\rho}(\vec{y})\hat{H}_0\hat{\rho}(\vec{x})-\hat{\rho}(\vec{y})\hat{\rho}(\vec{x})\hat{H}_0$$

$$\tag{7.9}$$

Sandwich the above expression between the state vector $\langle 0|$ and its dual $|0\rangle$ and use

$\hat{H}_0|0\rangle=0$ and $\langle 0|\hat{H}_0=0$ to obtain,

$$i\vec{\nabla}_{\vec{x}}\cdot\langle 0|\left[\hat{\rho}(\vec{y}),\hat{\vec{J}}(\vec{x})\right]|0\rangle=\langle 0|\hat{\rho}(\vec{x})\hat{H}_0\hat{\rho}(\vec{y})|0\rangle+\langle 0|\hat{\rho}(\vec{y})\hat{H}_0\hat{\rho}(\vec{x})|0\rangle \tag{7.10}$$

Next set $\vec{y}=\vec{x}$ to obtain,

$$i\vec{\nabla}_{\vec{x}}\cdot\langle 0|\left[\hat{\rho}(\vec{y}),\hat{\vec{J}}(\vec{x})\right]|0\rangle\Big|_{\vec{y}=\vec{x}}=2\langle 0|\hat{\rho}(\vec{x})\hat{H}_0\hat{\rho}(\vec{x})|0\rangle \tag{7.11}$$

Use (5.16) in the above to obtain,

$$i\vec{\nabla}_{\vec{x}}\cdot\langle 0|\left[\hat{\rho}(\vec{y}),\hat{\vec{J}}(\vec{x})\right]|0\rangle\Big|_{\vec{y}=\vec{x}}=2\sum_{|n\rangle,|m\rangle}\langle 0|\hat{\rho}(\vec{x})|n\rangle\langle n|\hat{H}_0|m\rangle\langle m|\hat{\rho}(\vec{x})|0\rangle \tag{7.12}$$

Next use (5.16) and (5.13) to obtain,

$$i\vec{\nabla}_{\vec{x}}\cdot\langle 0|\left[\hat{\rho}(\vec{y}),\hat{\vec{J}}(\vec{x})\right]|0\rangle\Big|_{\vec{y}=\vec{x}}=2\sum_{|n\rangle}\xi_{|n\rangle}\langle 0|\hat{\rho}(\vec{x})|n\rangle\langle n|\hat{\rho}(\vec{x})|0\rangle=2\sum_{|n\rangle}\xi_{|n\rangle}\left|\langle 0|\hat{\rho}(\vec{x})|n\rangle\right|^2$$

$$\tag{7.13}$$

Now, in general, the quantity $\langle 0|\hat{\rho}(\vec{x})|n\rangle$ is not zero [13] and since $\xi_{|n\rangle}>0$ for all

$|n\rangle\neq|0\rangle$ the above expression is non-zero and positive. Therefore the Schwinger term

must be non-zero. This is, of course, in direct contradiction to (7.4).

As a result of the above discussion we see that the various elements of quantum field theory do not produce consistent results. On one hand if the continuity equation is true then the Schwinger term must be zero. On the other hand if the vacuum state $|0\rangle$ is a lower bound to the free field energy (per Eqs. (5.15) or (5.20)) then the Schwinger term cannot be zero.

### VIII. Evaluating the Schwinger term

To confirm the results of the previous section we will use the material of section V to show that the Schwinger term is, indeed, nonzero. Define,

$$\vec{I}(\vec{y},\vec{x}) = \langle 0|\left[\hat{\rho}(\vec{y}),\hat{\vec{J}}(\vec{x})\right]|0\rangle = \langle 0|ST(\vec{y},\vec{x})|0\rangle \qquad (8.1)$$

If $ST(\vec{y},\vec{x})$ is zero then $\vec{I}(\vec{y},\vec{x})$ must be zero. Use (4.18) in the above to yield,

$$\vec{I}(\vec{y},\vec{x}) = q^2\langle 0|\left[\hat{\psi}^\dagger(\vec{y})\hat{\psi}(\vec{y}),\hat{\psi}^\dagger(\vec{x})\vec{\alpha}\hat{\psi}(\vec{x})\right]|0\rangle \qquad (8.2)$$

Use (5.5) and (5.6) in the above to obtain,

$$\vec{I}(\vec{y},\vec{x}) = q^2\sum_{n,m,r,s=1}^{\infty}\langle 0|\left[\begin{pmatrix}\hat{b}_n^\dagger\varphi_n^{(0)\dagger}(\vec{y})\\+\hat{d}_n\varphi_{-n}^{(0)\dagger}(\vec{y})\end{pmatrix}\begin{pmatrix}\hat{b}_m\varphi_m^{(0)}(\vec{y})\\+\hat{d}_m^\dagger\varphi_{-m}^{(0)}(\vec{y})\end{pmatrix},\\\begin{pmatrix}\hat{b}_r^\dagger\varphi_r^{(0)\dagger}(\vec{x})\\+\hat{d}_r\varphi_{-r}^{(0)\dagger}(\vec{x})\end{pmatrix}\vec{\alpha}\begin{pmatrix}\hat{b}_s\varphi_s^{(0)}(\vec{x})\\+\hat{d}_s^\dagger\varphi_{-s}^{(0)}(\vec{x})\end{pmatrix}\right]|0\rangle \qquad (8.3)$$

Use (5.8) and (5.9) in the above to yield,

$$\vec{I}(\vec{y},\vec{x}) = q^2\sum_{\substack{n,m,\\r,s=1}}^{\infty}\begin{pmatrix}\langle 0|\hat{b}_n^\dagger\hat{d}_m^\dagger,\hat{d}_r\hat{b}_s|0\rangle\left(\varphi_n^{(0)\dagger}(\vec{y})\varphi_{-m}^{(0)}(\vec{y})\right)\left(\varphi_{-r}^{(0)\dagger}(\vec{x})\vec{\alpha}\varphi_s^{(0)}(\vec{x})\right)\\+\langle 0|\hat{d}_n\hat{b}_m,\hat{b}_r^\dagger\hat{d}_s^\dagger|0\rangle\left(\varphi_{-n}^{(0)\dagger}(\vec{y})\varphi_m^{(0)}(\vec{y})\right)\left(\varphi_r^{(0)\dagger}(\vec{x})\vec{\alpha}\varphi_s^{(0)}(\vec{x})\right)\\+\langle 0|\hat{d}_n\hat{d}_m^\dagger,\hat{d}_r\hat{d}_s^\dagger|0\rangle\left(\varphi_{-n}^{(0)\dagger}(\vec{y})\varphi_{-m}^{(0)}(\vec{y})\right)\left(\varphi_{-r}^{(0)\dagger}(\vec{x})\vec{\alpha}\varphi_{-s}^{(0)}(\vec{x})\right)\end{pmatrix} \qquad (8.4)$$

From (5.8) and (5.9) we have the following relationships,

$$\langle 0|\hat{b}_n^\dagger \hat{d}_m^\dagger, \hat{d}_r \hat{b}_s |0\rangle = \langle 0|\hat{b}_n^\dagger \hat{d}_m^\dagger \hat{d}_r \hat{b}_s |0\rangle - \langle 0|\hat{d}_r \hat{b}_s \hat{b}_n^\dagger \hat{d}_m^\dagger |0\rangle = -\delta_{mr}\delta_{ns}$$

$$\langle 0|\hat{d}_n \hat{b}_m, \hat{b}_r^\dagger \hat{d}_s^\dagger |0\rangle = \langle 0|\hat{d}_n \hat{b}_m \hat{b}_r^\dagger \hat{d}_s^\dagger |0\rangle - \langle 0|\hat{b}_r^\dagger \hat{d}_s^\dagger \hat{d}_n \hat{b}_m |0\rangle = \delta_{sn}\delta_{rm}$$

$$\langle 0|\hat{d}_n \hat{d}_m^\dagger, \hat{d}_r \hat{d}_s^\dagger |0\rangle = \langle 0|\hat{d}_n \hat{d}_m^\dagger \hat{d}_r \hat{d}_s^\dagger |0\rangle - \langle 0|\hat{d}_r \hat{d}_s^\dagger \hat{d}_n \hat{d}_m^\dagger |0\rangle = \delta_{sr}\delta_{mn} - \delta_{sr}\delta_{mn} = 0$$

Use these in (8.4) to obtain,

$$\vec{I}(\vec{y},\vec{x}) = q^2 \sum_{n,m=1}^{\infty} \begin{pmatrix} \left(\varphi_{-n}^{(0)\dagger}(\vec{y})\varphi_m^{(0)}(\vec{y})\right)\left(\varphi_m^{(0)\dagger}(\vec{x})\vec{\alpha}\varphi_{-n}^{(0)}(\vec{x})\right) \\ -\left(\varphi_n^{(0)\dagger}(\vec{y})\varphi_{-m}^{(0)}(\vec{y})\right)\left(\varphi_{-m}^{(0)\dagger}(\vec{x})\vec{\alpha}\varphi_n^{(0)}(\vec{x})\right) \end{pmatrix} \tag{8.5}$$

By relabeling some of the dummy indices this can be rewritten as,

$$\vec{I}(\vec{y},\vec{x}) = \left\{ q^2 \sum_{n,m=1}^{\infty} \left(\varphi_{-n}^{(0)\dagger}(\vec{y})\varphi_m^{(0)}(\vec{y})\right)\left(\varphi_m^{(0)\dagger}(\vec{x})\vec{\alpha}\varphi_{-n}^{(0)}(\vec{x})\right) \right\} - (c.c.) \tag{8.6}$$

where the expression c.c. means to take the complex conjugate of the previous term.

Next take the divergence of the above expression with respect to $\vec{x}$ to obtain,

$$\vec{\nabla}_{\vec{x}} \cdot \vec{I}(\vec{y},\vec{x}) = \left\{ q^2 \sum_{n,m=1}^{\infty} \left(\varphi_{-n}^{(0)\dagger}(\vec{y})\varphi_m^{(0)}(\vec{y})\right)\vec{\nabla}\cdot\left(\varphi_m^{(0)\dagger}(\vec{x})\vec{\alpha}\varphi_{-n}^{(0)}(\vec{x})\right) \right\} - (c.c.) \tag{8.7}$$

To evaluate this expression use the following,

$$\vec{\nabla}\cdot\left(\varphi_m^{(0)\dagger}(\vec{x})\vec{\alpha}\varphi_{-n}^{(0)}(\vec{x})\right) = \left(\left(\vec{\alpha}\cdot\vec{\nabla}\varphi_m^{(0)}(\vec{x})\right)^\dagger \varphi_{-n}^{(0)}(\vec{x})\right) + \left(\varphi_m^{(0)\dagger}(\vec{x})\vec{\alpha}\cdot\vec{\nabla}\varphi_{-n}^{(0)}(\vec{x})\right) \tag{8.8}$$

Next use (2.3) in the above to obtain,

$$\vec{\nabla}\cdot\left(\varphi_m^{(0)\dagger}(\vec{x})\vec{\alpha}\varphi_{-n}^{(0)}(\vec{x})\right) = \left(\left(iH_0\varphi_m^{(0)}(\vec{x})\right)^\dagger \varphi_{-n}^{(0)}(\vec{x})\right) + \left(\varphi_m^{(0)\dagger}(\vec{x})iH_0\varphi_{-n}^{(0)}(\vec{x})\right) \tag{8.9}$$

Next use (2.15) to obtain,

$$\vec{\nabla}\cdot\left(\varphi_m^{(0)\dagger}(\vec{x})\vec{\alpha}\varphi_{-n}^{(0)}(\vec{x})\right) = i\varphi_m^{(0)\dagger}(\vec{x})\left(\lambda_{-n}E_{-n} - \lambda_m E_m\right)\varphi_{-n}^{(0)}(\vec{x}) \tag{8.10}$$

Use this in (8.7) to yield,

$$\vec{\nabla}_{\vec{x}} \cdot \vec{I}\left(\vec{y}, \vec{x}\right) = \left\{ iq^2 \sum_{n,m=1}^{\infty} \left(\varphi_{-n}^{(0)\dagger}\left(\vec{y}\right) \varphi_m^{(0)}\left(\vec{y}\right)\right) \left(\varphi_m^{(0)\dagger}\left(\vec{x}\right) \varphi_{-n}^{(0)}\left(\vec{x}\right)\right) \begin{pmatrix} \lambda_{-n} E_{-n} \\ -\lambda_m E_m \end{pmatrix} \right\} - \left(c.c.\right)$$

(8.11)

Use the fact that, in the above expression, $\lambda_{-n} = -1$, $\lambda_m = 1$, $E_{-n} = E_n$ and set $\vec{y} = \vec{x}$ to obtain,

$$\vec{\nabla}_{\vec{x}} \cdot \vec{I}\left(\vec{y}, \vec{x}\right)\Big|_{\vec{y}=\vec{x}} = -2\left\{ iq^2 \sum_{n,m=1}^{\infty} \left(\varphi_{-n}^{(0)\dagger}\left(\vec{x}\right) \varphi_m^{(0)}\left(\vec{x}\right)\right) \left(\varphi_m^{(0)\dagger}\left(\vec{x}\right) \varphi_{-n}^{(0)}\left(\vec{x}\right)\right) \left(E_n + E_m\right) \right\}$$    (8.12)

This yields,

$$\vec{\nabla}_{\vec{x}} \cdot \vec{I}\left(\vec{y}, \vec{x}\right)\Big|_{\vec{y}=\vec{x}} = -2\left\{ iq^2 \sum_{n,m=1}^{\infty} \left|\left(\varphi_m^{(0)\dagger}\left(\vec{x}\right) \varphi_{-n}^{(0)}\left(\vec{x}\right)\right)\right|^2 \left(E_n + E_m\right) \right\}$$

(8.13)

Each term in the sum is positive so that the above expression is not equal to zero. This confirms that the Schwinger term $ST\left(\vec{y}, \vec{x}\right)$ is non-zero. This result is expected because, as previously discussed, the Schwinger term must be non-zero if (5.15) is true.

## IX.  The Schwinger term and Gauge invariance

In Section VII it was shown that the Schwinger term must be zero for the continuity equation to be true for arbitrary values of the electric potential.   In this section we will also show that the Schwinger term must be zero for the theory to be gauge invariant.  To demonstrate this, take the time derivative of the current expectation value (6.4) to yield,

$$\frac{\partial \vec{J}_e\left(\vec{x},t\right)}{\partial t} = i\left\langle \Omega(t)\left| \left[ \hat{H}(t), \hat{\vec{J}}\left(\vec{x}\right) \right] \right| \Omega(t) \right\rangle$$

(9.1)

Use (4.22) in the above to obtain,

$$\frac{\partial \vec{J}_e\left(\vec{x},t\right)}{\partial t} = i\left\langle\Omega(t)\left|\begin{pmatrix}\left[\hat{H}_0,\hat{\vec{J}}\left(\vec{x}\right)\right]-\int\left[\hat{\vec{J}}\left(\vec{y}\right)\cdot\vec{A}\left(\vec{y},t\right),\hat{\vec{J}}\left(\vec{x}\right)\right]d\vec{y}\\+\int\left[\hat{\rho}\left(\vec{y}\right),\hat{\vec{J}}\left(\vec{x}\right)\right]A_0\left(\vec{y},t\right)d\vec{y}\end{pmatrix}\right|\Omega(t)\right\rangle \qquad (9.2)$$

Next perform the gauge transformation (2.9) to obtain,

$$\frac{\partial \vec{J}_e\left(\vec{x},t\right)}{\partial t} = i\left\langle\Omega(t)\left|\begin{pmatrix}\left[\hat{H}_0,\hat{\vec{J}}\left(\vec{x}\right)\right]-\int\left[\hat{\vec{J}}\left(\vec{y}\right)\cdot\left(\vec{A}\left(\vec{y},t\right)-\vec{\nabla}\chi\left(\vec{y},t\right)\right),\hat{\vec{J}}\left(\vec{x}\right)\right]d\vec{y}\\+\int\left[\hat{\rho}\left(\vec{y}\right),\hat{\vec{J}}\left(\vec{x}\right)\right]\left(A_0\left(\vec{y},t\right)+\frac{\partial\chi\left(\vec{y},t\right)}{\partial t}\right)d\vec{y}\end{pmatrix}\right|\Omega(t)\right\rangle$$

$$(9.3)$$

The quantity $\partial\vec{J}_e/\partial t$ is a physical observable and therefore, if the theory is gauge invariant, must not depend on the quantities $\chi$ or $\partial\chi/\partial t$. Now, at a particular instant of time $\partial\chi/\partial t$ can be varied in an arbitrary manner without changing the values, at that instant of time, of any of the other quantities on the right hand side of the equals sign in the above equation. Therefore for $\partial\vec{J}_e/\partial t$ to be independent of $\partial\chi/\partial t$ we must have that $ST\left(\vec{y},\vec{x}\right)=\left[\hat{\rho}\left(\vec{y}\right),\hat{\vec{J}}\left(\vec{x}\right)\right]=0$. However, we have just shown that the Schwinger term is nonzero therefore we expect that there should be a problem with the gauge invariance of Dirac field theory.

This problem shows up when the vacuum polarization tensor $\pi^{\mu\nu}$ is calculated. The first order change in the vacuum current due to an applied electric potential can be shown to be,

$$J_{vac}^\mu\left(\vec{x},t\right)=\int\pi^{\mu\nu}\left(\vec{x}-\vec{x}',t-t',\right)A_\nu\left(\vec{x}',t'\right)d\vec{x}'dt' \qquad (9.4)$$

It is well known that when the vacuum polarization tensor, $\pi^{\mu\nu}$, is calculated, using perturbation theory, the result is not gauge invariant (see Chapter 14 of [3], Sect. 22 of [6], Chapter 5 of [14], and [15]). The non-gauge invariant terms must be removed from the results of the calculation in order to obtain a physically correct result. This may involve some form of regularization, where other functions are introduced that happen to have the correct behavior so that the non-gauge invariant terms are cancelled. However there is no physical explanation for introducing these functions [15]. They are merely mathematical devices used to force the desired gauge invariant result.

Consider, for example, a calculation of the vacuum polarization tensor by Heitler (see page 322 of ref. 6). Heitler's solution for the Fourier transform of the vacuum polarization tensor is given by

$$\pi^{uv}\left(k^{\alpha}\right) = \pi_{G}^{uv}\left(k^{\alpha}\right) + \pi_{NG}^{uv}\left(k^{\alpha}\right) \tag{9.5}$$

where $k^{\alpha}$ is the 4 momentum of the electromagnetic field. The first term on the right hand side is given by,

$$\pi_{G}^{uv}\left(k^{\alpha}\right) = \left(\frac{2q^2}{3\pi}\right)\left(k^{\mu}k^{\nu} - g^{\mu\nu}k^2\right)\int_{2m}^{\infty}dz\,\frac{\left(z^2 + 2m^2\right)\sqrt{\left(z^2 - 4m^2\right)}}{z^2\left(z^2 - k^2\right)} \tag{9.6}$$

This term is gauge invariant because $k_{\nu}\pi_{G}^{uv} = 0$.

The second term on the right of (9.5) is

$$\pi_{NG}^{uv}\left(k^{\alpha}\right) = \left(\frac{2q^2}{3\pi}\right)g_{\nu}^{\mu}\left(1 - g^{\mu 0}\right)\int_{2m}^{\infty}dz\,\frac{\left(z^2 + 2m^2\right)\sqrt{\left(z^2 - 4m^2\right)}}{z^2} \tag{9.7}$$

where there is no summation over the two $\mu$ superscripts that appear on the right. Note that $\pi_{NG}^{uv}$ is not gauge invariant because $k_{\nu}\pi_{NG}^{uv} \neq 0$. Therefore to get a physically correct result it is necessary to "correct" equation (9.5) by dropping $\pi_{NG}^{uv}$ from the solution.

The problem that we are addressing is why does this extra non-gauge invariant term appear in a theory that is supposed to be gauge invariant? In order to understand the source of the problem we shall use perturbation theory to calculate the vacuum current and show that the lack of gauge invariance of the result is directly related to the fact the Schwinger term is non-zero. This problem was originally addressed in [12]. The following analysis closely follows this reference.

The first order change in the vacuum current due to an applied electric potential is given by the following expression which is derived in Appendix A (see also [12] and Eq. 8.3 of [16]).

$$\vec{J}_{vac}\left(\vec{x},t\right) = i\left\langle 0\left|\left[\hat{\vec{J}}_I\left(\vec{x},t\right),\int d\vec{y}\int_{-\infty}^{t}dt'\left(-\hat{\vec{J}}_I\left(\vec{y},t'\right)\cdot\vec{A}\left(\vec{y},t'\right)+\hat{\rho}_I\left(\vec{y},t'\right)A_o\left(\vec{y},t'\right)\right)\right]\right|0\right\rangle \quad (9.8)$$

where the operators $\hat{\vec{J}}_I\left(\vec{x},t\right)$ and $\hat{\rho}_I\left(\vec{x},t\right)$ are the current and charge operators, respectively, in the interaction representation. They are defined by,

$$\hat{\vec{J}}_I\left(\vec{x},t\right) = e^{i\hat{H}_0 t}\hat{\vec{J}}\left(\vec{x}\right)e^{-i\hat{H}_0 t} \text{ and } \hat{\rho}_I\left(\vec{x},t\right) = e^{i\hat{H}_0 t}\hat{\rho}\left(\vec{x}\right)e^{-i\hat{H}_0 t} \quad (9.9)$$

They can be shown to satisfy the continuity equation (see Eq. 3.11 of [16] ),

$$\frac{\partial\hat{\rho}_I\left(\vec{x},t\right)}{\partial t} = -\vec{\nabla}\cdot\hat{\vec{J}}_I\left(\vec{x},t\right) \quad (9.10)$$

The change in the vacuum current $\delta_g\vec{J}_{vac}\left(\vec{x},t\right)$ due to a gauge transformation is obtained by using (2.9) in (9.8) to yield

$$\delta_g \vec{J}_{vac}(\vec{x},t) = i \left\langle 0 \left\| \hat{\vec{J}}_I(\vec{x},t), \int d\vec{y} \int_{-\infty}^{t} dt' \left( \hat{\vec{J}}_I(\vec{y},t') \cdot \vec{\nabla} \chi(\vec{y},t') + \hat{\rho}_I(\vec{y},t') \frac{\partial \chi(\vec{y},t')}{\partial t'} \right) \right\| 0 \right\rangle$$

(9.11)

If quantum field theory is gauge invariant then a gauge transformation should produce no change in any observable quantity. Therefore $\delta_g \vec{J}_{vac}(\vec{x},t)$ should be zero. To see if this is the case we will solve the above equation as follows. First consider the following relationship,

$$\int_{-\infty}^{t} dt' \hat{\rho}_I(\vec{y},t') \frac{\partial \chi(\vec{y},t')}{\partial t'} = \Big|_{-\infty}^{t} \hat{\rho}_I(\vec{y},t') \chi(\vec{y},t') - \int_{-\infty}^{t} dt' \chi(\vec{y},t') \frac{\partial \hat{\rho}_I(\vec{y},t')}{\partial t'} \qquad (9.12)$$

Assume that $\chi(\vec{y},t) = 0$ at $t \to -\infty$. Use this and (9.10) in the above expression to obtain

$$\int_{-\infty}^{t} dt' \hat{\rho}_I(\vec{y},t') \frac{\partial \chi(\vec{y},t')}{\partial t'} = \hat{\rho}_I(\vec{y},t) \chi(\vec{y},t) + \int_{-\infty}^{t} dt' \chi(\vec{y},t') \vec{\nabla} \cdot \vec{J}_I(\vec{y},t') \qquad (9.13)$$

Substitute this into (9.11) to obtain

$$\delta_g \vec{J}_{vac}(\vec{x},t) = i \left\langle 0 \left\| \hat{\vec{J}}_I(\vec{x},t), \int d\vec{y} \int_{-\infty}^{t} dt' \left( \hat{\vec{J}}_I(\vec{y},t') \cdot \vec{\nabla} \chi(\vec{y},t') + \chi(\vec{y},t') \vec{\nabla} \cdot \hat{\vec{J}}_I(\vec{y},t') \right) \right\| 0 \right\rangle$$
$$+ i \left\langle 0 \left\| \hat{\vec{J}}_I(\vec{x},t), \int \hat{\rho}_I(\vec{y},t) \chi(\vec{y},t) d\vec{y} \right\| 0 \right\rangle$$

(9.14)

Rearrange terms to obtain

$$\delta_g \vec{J}_{vac}(\vec{x},t) = i \left\langle 0 \left\| \hat{\vec{J}}_I(\vec{x},t), \int_{-\infty}^{t} dt' \int d\vec{y} \vec{\nabla} \cdot \left( \hat{\vec{J}}_I(\vec{y},t') \chi(\vec{y},t') \right) \right\| 0 \right\rangle$$
$$+ i \left\langle 0 \left\| \hat{\vec{J}}_I(\vec{x},t), \int \hat{\rho}_I(\vec{y},t) \chi(\vec{y},t) d\vec{y} \right\| 0 \right\rangle$$

(9.15)

Assume reasonable boundary conditions at $|\vec{y}| \to \infty$ so that

$$\int d\vec{y}\vec{\nabla}\cdot\left(\hat{\vec{J}}_I\left(\vec{y},t'\right)\chi\left(\vec{y},t'\right)\right)=0 \tag{9.16}$$

Use this to obtain

$$\begin{aligned}\delta_g\vec{J}_{vac}\left(\vec{x},t\right)&=i\left\langle 0\left|\left[\hat{\vec{J}}_I\left(\vec{x},t\right),\int\hat{\rho}_I\left(\vec{y},t\right)\chi\left(\vec{y},t\right)d\vec{y}\right]\right|0\right\rangle\\&=i\int\left\langle 0\left|\left[\hat{\vec{J}}_I\left(\vec{x},t\right),\hat{\rho}_I\left(\vec{y},t\right)\right]\right|0\right\rangle\chi\left(\vec{y},t\right)d\vec{y}\end{aligned} \tag{9.17}$$

Use (9.9) and the fact that $H_0\left|0\right\rangle=\left\langle 0\right|H_0=0$ to show that,

$$\left\langle 0\left|\left[\hat{\vec{J}}_I\left(\vec{x},t\right),\hat{\rho}_I\left(\vec{y},t\right)\right]\right|0\right\rangle=\left\langle 0\left|\left[\hat{\vec{J}}\left(\vec{x}\right),\hat{\rho}\left(\vec{y}\right)\right]\right|0\right\rangle=-\left\langle 0\left|ST\left(\vec{y},\vec{x}\right)\right|0\right\rangle \tag{9.18}$$

Use this in (9.17) to obtain,

$$\delta_g\vec{J}_{vac}\left(\vec{x},t\right)=-i\int\left\langle 0\left|ST\left(\vec{y},\vec{x}\right)\right|0\right\rangle\chi\left(\vec{y},t\right)d\vec{y} \tag{9.19}$$

Therefore for $\delta_g\vec{J}_{vac}\left(\vec{x},t\right)$ to be zero, for arbitrary $\chi\left(\vec{y},t\right)$, the quantity $\left\langle 0\left|ST\left(\vec{x},\vec{y}\right)\right|0\right\rangle$ must be zero. However, we have shown in Section VIII that this quantity is not zero. Therefore perturbation theory does not produce a gauge invariant result for the vacuum current. This is why regulation is required. It is needed to remove the non-gauge invariant terms that occur due to the fact that the Schwinger term is nonzero.

## X. Removing energy from the vacuum state in hole theory

In the last several sections we have examined the problems that quantum field theory has with gauge invariance and the continuity equation. We have shown that these problems are related to the fact that in quantum field theory the vacuum state $\left|0\right\rangle$ is the state with the lowest free field energy. In this section we will examine the vacuum state in hole theory to determine if it is the state with the lowest free field energy. It will be shown that this is not the case. That is, in hole theory there exist states with less free field energy than the vacuum state.

Recall that in hole theory the vacuum state corresponds to the state in which every negative energy state is filled up with a single electron and all the positive energy states are unoccupied. Each of these negative energy electrons obeys the single particle Dirac equation. If an external electric field is applied for a finite amount of time then the energy of each negative energy electron will change in response to the electric field. The total change in the energy of the vacuum state is the sum of the change in the energy of each individual electron. It will be shown that it is possible to find an electric field where this change is negative. That is, energy is extracted from the initial vacuum state due to its interaction with the electric field.

To show this we will work the following problem. Assume that at some initial time $t = t_i$ the system is in the unperturbed vacuum state. This is the state where all the negative energy states $\varphi_{-j}^{(0)}(\vec{x}, t_i)$ (where j is positive) are occupied by a single electron. (Note that we are using the ordering convention introduced in the paragraph leading up to equation (5.1)). Using the notation of Section III the vacuum state can be written as,

$$\psi^N(\vec{x}_1, \vec{x}_2, ..., \vec{x}_N, t_i) \underset{N \to \infty}{=} \frac{1}{\sqrt{N!}} \sum_P (-1)^P P\left(\varphi_{-1}^{(0)}(\vec{x}_1, t_i)\varphi_{-2}^{(0)}(\vec{x}_2, t_i)\cdots\varphi_{-N}^{(0)}(\vec{x}_N, t_i)\right) (10.1)$$

Next, at $t = t_i$, apply an electric potential which is removed at some later time $t = t_f$. Under the action of the electric potential each initial state $\varphi_{-j}^{(0)}(\vec{x}, t_i)$ evolves into the final state $\varphi_{-j}(\vec{x}, t_f)$. Therefore the perturbed vacuum state, at time $t = t_f$, is,

$$\psi^N(\vec{x}_1, \vec{x}_2, ..., \vec{x}_N, t_f) \underset{N \to \infty}{=} \frac{1}{\sqrt{N!}} \sum_P (-1)^P P\left(\varphi_{-1}(\vec{x}_1, t_f)\varphi_{-2}(\vec{x}_2, t_f)\cdots\varphi_{-N}(\vec{x}_N, t_f)\right)$$

$$(10.2)$$

To simplify notation define,

$$\langle f(\vec{x}) \rangle = \int f(\vec{x}) d\vec{x} \tag{10.3}$$

The total change in the free field energy of the vacuum state, $\Delta \xi_{f,vac}(t_i \to t_f)$, is the sum of changes of the free field energy of each individual vacuum electron. Therefore,

$$\Delta \xi_{f,vac}(t_i \to t_f)_{N \to \infty} = \sum_{j=1}^{N} \Delta \xi_{f,-j}(t_i \to t_f) \tag{10.4}$$

where $\Delta \xi_{f,n}(t_i \to t_f)$ is the change in the free field energy of the electron in state 'n'.

Since the initial state for this electron is $\varphi_n^{(0)}(\vec{x}, t_i)$ and the final state is $\varphi_n(\vec{x}, t_f)$,

$$\Delta \xi_{f,n}(t_i \to t_f) = \langle \varphi_n^{\dagger}(\vec{x}, t_f) H_0 \varphi_n(\vec{x}, t_f) \rangle - \langle \varphi_n^{(0)\dagger}(\vec{x}, t_i) H_0 \varphi_n^{(0)}(\vec{x}, t_i) \rangle \tag{10.5}$$

Now we want to determine $\Delta \xi_{f,vac}(t_i \to t_f)$ for a given electric potential applied during the time interval between $t_i$ and $t_f$. To do this it is necessary to evaluate $\varphi_n(\vec{x}, t)$. The change in $\varphi_n(\vec{x}, t)$, due to an applied electric potential, must be determined using perturbation theory since it is, in general, not possible to find exact solutions to these kind of problems. This problem has already been considered for the case of field theory in Appendix A. Exactly the same analysis can be applied for the single particle wave function. The result is that $\varphi_n(\vec{x}, t)$ can be expanded as a series in the charge 'q' i.e.,

$$\varphi_n(\vec{x}, t) = \varphi_n^{(0)}(\vec{x}, t) + q\varphi_n^{(1)}(\vec{x}, t) + q^2 \varphi_n^{(2)}(\vec{x}, t) + O(q^3) \tag{10.6}$$

where $O(q^3)$ means terms to the third order in the perturbation or higher. Use this in (10.5) to obtain $\Delta \xi_{f,n}(t_i \to t_f)$. It is shown in Appendix B that

$$\Delta \xi_{f,n}(t_i \to t_f) = \xi_f(\varphi_n(\vec{x}, t_f)) - \lambda_n E_n = q^2 \left( \langle \varphi_n^{(1)\dagger} H_0 \varphi_n^{(1)} \rangle - \lambda_n E_n \langle \varphi_n^{(1)\dagger} \varphi_n^{(1)} \rangle \right) + O(q^3)$$

$$\tag{10.7}$$

where,

$$\varphi_n^{(1)}\left(\vec{x},t\right) = -i \int_{-\infty}^{t} dt' e^{-iH_0(t-t')} V\left(\vec{x},t'\right) \varphi_n^{(0)}\left(\vec{x},t'\right) \tag{10.8}$$

and $V = \left(-\vec{\alpha}\cdot\vec{A} + A_0\right)$. This result can be used in (10.4) to obtain $\Delta\xi_{f,vac}\left(t_i \to t_f\right)$ to the second order in q.

Now, in order to avoid unnecessary mathematical details, this problem will be worked in 1-1D space-time where we will take the space axis in the z-direction. In 1-1D space-time the Dirac equation can be written as,

$$i\frac{\partial\psi}{\partial t} = \left(\sigma_x\left(-i\frac{\partial}{\partial z} - qA_z\right) + m\sigma_z + qA_0\right)\psi \tag{10.9}$$

where $\sigma_x$ and $\sigma_z$ are the Pauli spin matrices,

$$\sigma_x = \begin{pmatrix} 0 & 1 \\ 1 & 0 \end{pmatrix} \quad \text{and} \quad \sigma_z = \begin{pmatrix} 1 & 0 \\ 0 & -1 \end{pmatrix} \tag{10.10}$$

The orthonormal free field solutions (the electric potential is zero) of (10.9) are given by,

$$\varphi_{\lambda,p}^{(0)}\left(z,t\right) = u_{\lambda,p} e^{-i(\lambda E t - pz)} \tag{10.11}$$

where,

$$u_{\lambda,p} = N_{\lambda,p}\begin{pmatrix} 1 \\ p/(\lambda E + m) \end{pmatrix}; \quad N_{\lambda,p} = \sqrt{\frac{\lambda E + m}{2L\lambda E}}; \quad E = \sqrt{p^2 + m^2} \tag{10.12}$$

In the above expressions $L \to \infty$ is the 1 dimensional integration volume, p is the momentum, $\lambda = 1$ for positive energy states, and $\lambda = -1$ for negative energy states. The function $\varphi_{\lambda,p}^{(0)}\left(z,t\right)$ shall satisfy periodic boundary conditions $\varphi_{\lambda,p}^{(0)}\left(z,t\right) = \varphi_{\lambda,p}^{(0)}\left(z+L,t\right)$. This condition is realized if the momentum is given by,

$$p_r = \frac{2\pi r}{L} \text{ where r is an integer.} \tag{10.13}$$

Let the electric potential be,

$$A_z = 0 \text{ and } A_0 = 4\cos(kz)\left(\frac{\sin(mt)}{t}\right) = \left(e^{ikz} + e^{-ikz}\right)\int\limits_{-m}^{+m} e^{iqt}dq \tag{10.14}$$

where m is the mass of the electron and $k < m$. It is obvious from the above expression that $A_0 \rightarrow 0$ at $t \rightarrow \pm\infty$. Under the action of this electric potential the each initial wave function $\varphi_{\lambda,p}^{(0)}(z, t_i)$, where $t_i \rightarrow -\infty$, evolves into the final wave function $\varphi_{\lambda,p}(z, t_f)$ where $t_f \rightarrow +\infty$. $\varphi_{\lambda,p}(z, t_f)$ can be expanded as a series in orders of 'q' as in (10.6).

Using perturbation theory it is shown in Appendix C that,

$$\varphi_{\lambda,p_r}^{(1)}(z, t_f) = -i\left\{\left(c_{\lambda,p_r,k}\varphi_{\lambda,p_r+k}^{(0)}(z, t_f)\right) + \left(c_{\lambda,p_r,-k}\varphi_{\lambda,p_r-k}^{(0)}(z, t_f)\right)\right\} \tag{10.15}$$

where,

$$c_{\lambda,p_r,k} = 2\pi L u_{\lambda,p_r+k}^{\dagger} u_{\lambda,p_r} \tag{10.16}$$

Use Eq. (10.15) and the fact that,

$$H_0\varphi_{\lambda',p_r}^{(0)} = \lambda' E_{p_r}\varphi_{\lambda',p_r}^{(0)} \text{ and } \left\langle \varphi_{\lambda,p_s}^{(0)\dagger}\varphi_{\lambda',p_r}^{(0)}\right\rangle = \delta_{sr}\delta_{\lambda'\lambda} \tag{10.17}$$

to obtain,

$$\left\langle \varphi_{\lambda,p_r}^{(1)\dagger} H_0 \varphi_{\lambda,p_r}^{(1)}\right\rangle = \left|c_{\lambda,p_r,k}\right|^2 \lambda E_{p_r+k} + \left|c_{\lambda,p_r,-k}\right|^2 \lambda E_{p_r-k} \tag{10.18}$$

and,

$$\left\langle \varphi_{\lambda,p_r}^{(1)\dagger} \varphi_{\lambda,p_r}^{(1)}\right\rangle = \left|c_{\lambda,p_r,k}\right|^2 + \left|c_{\lambda,p_r,-k}\right|^2 \tag{10.19}$$

Use the above in (10.7) to obtain,

$$\Delta \xi_f \left( \lambda, p_r; t_i \to t_f \right) = q^2 \left( \begin{array}{c} \left| c_{\lambda, p_r, k} \right|^2 \left( \lambda E_{p_r+k} - \lambda E_{p_r} \right) \\ + \left| c_{\lambda, p_r, -k} \right|^2 \left( \lambda E_{p_r-k} - \lambda E_{p_r} \right) \end{array} \right) + O\left( q^3 \right) \qquad (10.20)$$

where $\Delta \xi_f \left( \lambda, p_r; t_i \to t_f \right)$ is the change in free field energy of the initial state

$\varphi_{\lambda, p_r}^{(0)} \left( z, t_i \right)$ when it evolves into the final state $\varphi_{\lambda, p_r} \left( z, t_f \right)$ under the action of the electric potential. Use (10.16) and (10.12) the above to obtain,

$$\Delta \xi_f \left( \lambda, p_r; t_i \to t_f \right) = 2\pi^2 q^2 \lambda k \left( \frac{(p_r+k)}{E_{p_r+k}} - \frac{(p_r-k)}{E_{p_r-k}} \right) + O\left( q^3 \right) \qquad (10.21)$$

The details of the derivation of this equation are provided in Appendix D.

Now we are interested in the change of the free field energy of the electrons in negative energy states in which case $\lambda = -1$. Therefore, for negative energy states,

$$\Delta \xi_f \left( \lambda = -1, p_r; t_i \to t_f \right) = -2\pi^2 q^2 k \left( \frac{(p_r+k)}{E_{p_r+k}} - \frac{(p_r-k)}{E_{p_r-k}} \right) + O\left( q^3 \right) \qquad (10.22)$$

In the limit that $q \to 0$ the $O\left( q^3 \right)$ term can be dropped. Therefore,

$$\Delta \xi_f \left( \lambda = -1, p_r; t_i \to t_f \right)_{q \to 0} = -2\pi^2 q^2 k \left( \frac{(p_r+k)}{E_{p_r+k}} - \frac{(p_r-k)}{E_{p_r-k}} \right) \qquad (10.23)$$

It is shown in Appendix E that this quantity is negative for all $p_r$. Therefore the change in the free field energy of each vacuum electron is negative. Therefore the total change in the free field energy of the vacuum is negative. This means that, in hole theory, there exist quantum states with less energy than the vacuum state.

## XI. Discussion

We have just shown that in hole theory energy can be extracted from the vacuum state due to its interaction with an electric field. Now suppose we had worked the same problem in quantum field theory. From Appendix A the perturbed vacuum state $\left|0_p\right\rangle$, due to an interaction with an electromagnetic field, is given by,

$$\left|0_p\right\rangle = \left(1 - iqe^{-iH_0 t_f} \int_{t_i}^{t_f} \hat{V}_I(t)\,dt + \ldots\right)\left|0\right\rangle \tag{11.1}$$

Therefore, for an arbitrary perturbation, the final state $\left|0_p\right\rangle$ is the sum of the vacuum state $\left|0\right\rangle$ and various states of the form $\hat{b}_{j_1}^\dagger \hat{d}_{j_2}^\dagger \left|0\right\rangle$, $\hat{b}_{j_1}^\dagger \hat{b}_{j_2}^\dagger \hat{d}_{j_3}^\dagger \hat{d}_{j_4}^\dagger \left|0\right\rangle$, etc. As discussed

previously all these states are eigenstates of the free field Hamiltonian $\hat{H}_0$ with positive energy eigenvalues. Therefore the free field energy of $\left|0_p\right\rangle$ is positive. In field theory the effect of an interaction with the electric field can never result in the decrease in the free field energy of the vacuum state.

To understand why these differences between hole theory and field theory occur

consider the expression for the vacuum state in hole theory which is given by equation

(10.1). We will rewrite this using our notation for the 1-1D wave function where

$\varphi_n^{(0)} \to \varphi_{\lambda,p_r}^{(0)}$,

$$\Psi^{2N+1}(t_i) \underset{\substack{N\to\infty \\ \lambda=-1}}{=} \sum_P \frac{(-1)^p}{\sqrt{(2N+1)!}} P\begin{pmatrix} \varphi_{\lambda,p_0}^{(0)}(z_1,t_i)\,\varphi_{\lambda,p_1}^{(0)}(z_2,t_i)\,\varphi_{\lambda,p_{-1}}^{(0)}(z_3,t_i) \\ \cdots \varphi_{\lambda,p_N}^{(0)}(z_{2N},t_i)\,\varphi_{\lambda,p_{-N}}^{(0)}(z_{2N+1},t_i) \end{pmatrix} \tag{11.2}$$

In this expression the wave functions are ordered in terms of increasing magnitude of

momentum. Note that because of the way we have done the ordering the number of

electrons is 2N+1. Now as we have shown each of the initial unperturbed wave functions

$\varphi_{\lambda,p_r}^{(0)}(z,t_i)$ has evolved under the action of the electric potential into the final wave function $\varphi_{\lambda,p_r}(z,t_f)$. Therefore $\Psi^{2N+1}(t_i)$ evolves into,

$$\Psi^{2N+1}(t_f)_{\substack{N\to\infty \\ \lambda=-1}} = \sum_P \frac{(-1)^P}{\sqrt{(2N+1)!}} P\begin{pmatrix} \varphi_{\lambda,p_0}(z_1,t_f)\varphi_{\lambda,p_1}(z_2,t_f)\varphi_{\lambda,p_{-1}}(z_3,t_f) \\ \cdots \varphi_{\lambda,p_N}(z_{2N},t_f)\varphi_{\lambda,p_{-N}}(z_{2N+1},t_f) \end{pmatrix} \quad (11.3)$$

From the results of the previous section we have that to the lowest order,

$$\varphi_{\lambda,p_r}(z,t_f)_{q\to 0} = \varphi_{\lambda,p_r}^{(0)}(z,t_f) - iq\left\{\left(c_{\lambda,p_r,k}\phi_{\lambda,p_r+k}^{(0)}(z,t_f)\right)+\left(c_{\lambda,p_r,-k}\phi_{\lambda,p_r-k}^{(0)}(z,t_f)\right)\right\} \quad (11.4)$$

Next assume that $k = 2\pi s/L$ where 's' is a positive integer. We shall write $k_s$ instead of k to reflect this fact. In this case $p_r \pm k_s = p_{r\pm s}$ so that the above equation becomes,

$$\varphi_{\lambda,p_r}(z,t_f)_{q\to 0} = \varphi_{\lambda,p_r}^{(0)}(z,t_f) - iq\left\{\left(c_{\lambda,p_r,k_s}\varphi_{\lambda,p_{r+s}}^{(0)}(z,t_f)\right)+\left(c_{\lambda,p_r,k_{-s}}\varphi_{\lambda,p_{r-s}}^{(0)}(z,t_f)\right)\right\} \quad (11.5)$$

Use this in (11.3) to obtain,

$$\Psi^{2N+1}\left(t_f\right) \underset{\substack{N\to\infty \\ \lambda=-1 \\ q\to 0}}{=} \sum_P \frac{(-1)^p}{\sqrt{(2N+1)!}} P \begin{pmatrix} \left(\varphi_{\lambda,p_0}^{(0)}\left(z_1,t_f\right) - iq\left\{\begin{pmatrix} c_{\lambda,p_0,k_s}\phi_{\lambda,p_{0+s}}^{(0)}\left(z_1,t_f\right) \\ + \left(c_{\lambda,p_0,k_{-s}}\phi_{\lambda,p_{0-s}}^{(0)}\left(z_1,t_f\right)\right) \end{pmatrix}\right\}\right) \\ \dots \\ \left(\varphi_{\lambda,p_N}^{(0)}\left(z_{2N},t_f\right) - iq\left\{\begin{pmatrix} c_{\lambda,p_N,k_s}\phi_{\lambda,p_{N+s}}^{(0)}\left(z_{2N},t_f\right) \\ + \left(c_{\lambda,p_N,k_{-s}}\phi_{\lambda,p_{N-s}}^{(0)}\left(z_{2N},t_f\right)\right) \end{pmatrix}\right\}\right) \\ \left(\varphi_{\lambda,p_{-N}}^{(0)}\left(z_{2N+1},t_f\right) - iq\left\{\begin{pmatrix} c_{\lambda,p_{-N},k_s}\phi_{\lambda,p_{-N+s}}^{(0)}\left(z_{2N+1},t_f\right) \\ + \left(c_{\lambda,p_{-N},k_{-s}}\phi_{\lambda,p_{-N-s}}^{(0)}\left(z_{2N+1},t_f\right)\right) \end{pmatrix}\right\}\right) \end{pmatrix}$$

$$(11.6)$$

Now in the above expression we will only retain terms to the first order in q and write the result using the language of creation operators per section IV to obtain,

$$\left|\Psi^{2N+1}\left(t_f\right)\right\rangle \underset{\substack{N\to\infty \\ \lambda=-1 \\ q\to 0}}{=} \left(1 - iq\sum_{r=-N}^{N}\begin{pmatrix} e^{-i\lambda\left(E_{p_{r+s}}-E_{p_r}\right)t_f}c_{\lambda,p_r,k_s}\hat{a}_{\lambda,p_{r+s}}^{\dagger}\hat{a}_{\lambda,p_r} \\ + e^{-i\lambda\left(E_{p_{r-s}}-E_{p_r}\right)t_f}c_{\lambda,p_r,k_{-s}}\hat{a}_{\lambda,p_{r-s}}^{\dagger}\hat{a}_{\lambda,p_r} \end{pmatrix}\right)\left(e^{-iE_{vac}t_f}\left|\Psi_{vac}^{2N+1}\right\rangle\right)$$

$$(11.7)$$

where,

$$\left|\Psi_{vac}^{2N+1}\right\rangle \underset{\substack{N\to\infty \\ \lambda=-1}}{=} \left(\hat{a}_{\lambda,p_0}^{\dagger}\hat{a}_{\lambda,p_1}^{\dagger}\hat{a}_{\lambda,p_{-1}}^{\dagger}\hat{a}_{\lambda,p_2}^{\dagger}\hat{a}_{\lambda,p_{-2}}^{\dagger}\dots\hat{a}_{\lambda,p_N}^{\dagger}\hat{a}_{\lambda,p_{-N}}^{\dagger}\right)\Phi^0 \qquad (11.8)$$

and

$$E_{vac} = -\left(E_{p_0} + E_{p_1} + E_{p_{-1}} + \dots + E_{p_N} + E_{p_{-N}}\right) \qquad (11.9)$$

and where $\hat{a}_{\lambda,p_r}^{\dagger}$ $\left(\hat{a}_{\lambda,p_r}\right)$ creates (destroys) the state $\varphi_{\lambda,p_r}^{(0)}\left(z\right)$. The initial vacuum state (11.2) is given by,

$$\left|\Psi^{2N+1}\left(t_i\right)\right\rangle \underset{N\to\infty}{=} e^{-iE_{vac}t_i}\left|\Psi_{vac}^{2N+1}\right\rangle \qquad (11.10)$$

According to (11.8) $\left|\Psi_{vac}^{2N+1}\right\rangle$ is composed of the product of raising operators $\hat{a}_{-1,p_r}^\dagger$ where $-N \le r \le N$. The term $\hat{a}_{-1,p_{r+s}}^\dagger \hat{a}_{-1,p_r}$, in (11.7), acts on the state $\left|\Psi_{vac}^{2N+1}\right\rangle$ by removing the operator $\hat{a}_{-1,p_r}^\dagger$ from $\left|\Psi_{vac}^{2N+1}\right\rangle$ and replacing it with $\hat{a}_{-1,p_{r+s}}^\dagger$. The result will be zero if $\hat{a}_{-1,p_{r+s}}^\dagger$ is already one of the operators in the product of operators that make up $\left|\Psi_{vac}^{2N+1}\right\rangle$. This will be the case if $-N \le r+s \le N$. Therefore, since we assume 's' is positive, the term $\hat{a}_{-1,p_{r+s}}^\dagger \hat{a}_{-1,p_r}$ only produces a nonzero contribution when $r > N-s$. Similarly the term $\hat{a}_{-1,p_{r-s}}^\dagger \hat{a}_{-1,p_r}$ will only produce a nonzero contribution if $s-N > r$. Therefore we can write (11.7) as,

$$\left|\Psi^{2N+1}\left(t_f\right)\right\rangle \underset{q\to 0}{=} e^{-iE_{vac}t_f}\left(\Psi_{vac}^{2N+1} - iq\Psi_A^{2N+1} - iq\Psi_B^{2N+1}\right) \qquad (11.11)$$

where,

$$\left|\Psi_A^{2N+1}\right\rangle \underset{\substack{N\to\infty \\ \lambda=-1}}{=} \left(\sum_{r=N-s+1}^{N} e^{-i\lambda\left(E_{p_{r+s}}-E_{p_r}\right)t_f} c_{\lambda,p_r,k_s}\left(\hat{a}_{\lambda,p_{r+s}}^\dagger \hat{a}_{\lambda,p_r}\Psi_{vac}^{2N+1}\right)\right) \qquad (11.12)$$

and

$$\left|\Psi_B^{2N+1}\right\rangle \underset{\substack{N\to\infty \\ \lambda=-1}}{=} \left(\sum_{r=-N}^{r=s-N-1} e^{-i\lambda\left(E_{p_{r-s}}-E_{p_r}\right)t_f} c_{\lambda,p_r,k_{-s}}\left(\hat{a}_{\lambda,p_{r-s}}^\dagger \hat{a}_{\lambda,p_r}\Psi_{vac}^{2N+1}\right)\right) \qquad (11.13)$$

Now consider the terms $\hat{a}_{-1,p_{r+s}}^\dagger \hat{a}_{-1,p_r}\left|\Psi_{vac}^{2N+1}\right\rangle$ that appear in (11.12). The action of $\hat{a}_{-1,p_{r+s}}^\dagger \hat{a}_{-1,p_r}$ on $\left|\Psi_{vac}^{2N+1}\right\rangle$ is to destroy the state $\varphi_{-1,p_r}^{(0)}\left(z\right)$ with energy $-E_{p_r}$ and create the state $\varphi_{-1,p_{r+s}}^{(0)}(z)$ with energy $-E_{p_{r+s}}$ where $r+s > N$. Therefore the free field energy of the state $\hat{a}_{-1,p_{r+s}}^{\dagger}\,\hat{a}_{-1,p_r}\,\left|\Psi_{vac}^{2N+1}\right\rangle$ is given by,

$$\xi_f\left(\hat{a}_{-1,p_{r+s}}^{\dagger}\,\hat{a}_{-1,p_r}\,\left|\Psi_{vac}^{2N+1}\right\rangle\right) = E_{vac} - \left(E_{p_{r+s}} - E_{p_r}\right) < E_{vac} \text{ where } r+s > N \quad (11.14)$$

The term $\hat{a}_{-1,p_{r+s}}^{\dagger}\,\hat{a}_{-1,p_r}\,\left|\Psi_{vac}^{2N+1}\right\rangle$ is an eigenfunction of the free field Hamiltonian operator with an eigenvalue $E_{vac} - \left(E_{p_{r+s}} - E_{p_r}\right)$ which is less than $E_{vac}$, the energy of the vacuum state. Similarly the energy eigenvalue of the terms $\hat{a}_{-1,p_{r-s}}^{\dagger}\,\hat{a}_{-1,p_r}\,\left|\Psi_{vac}^{2N+1}\right\rangle$ that

appear in (11.13) is less than $E_{vac}$. Therefore the energy of the perturbed vacuum state

$\left|\Psi^{2N+1}\left(t_f\right)\right\rangle$ consists of a sum of states which include the original unperturbed vacuum

state $\left|\Psi_{vac}^{2N+1}\right\rangle$ and states of the form $\hat{a}_{-1,p_{r+s}}^{\dagger}\,\hat{a}_{-1,p_r}\,\left|\Psi_{vac}^{2N+1}\right\rangle$ and $\hat{a}_{-1,p_{r-s}}^{\dagger}\,\hat{a}_{-1,p_r}\,\left|\Psi_{vac}^{2N+1}\right\rangle$

that have less energy then the vacuum. The result is that the energy of the perturbed state

$\left|\Psi^{2N+1}\left(t_f\right)\right\rangle$ is less than that of the initial vacuum state $\left|\Psi^{2N+1}\left(t_i\right)\right\rangle$.

The reason for this result is that, in hole theory, we have defined the vacuum

$\left|\Psi_{vac}^{2N+1}\right\rangle$ using a limiting procedure. As defined in equations (11.2), (11.10), and (11.8)

the vacuum is the quantum state which consists of the product of terms of the form

$\hat{a}_{-1,p_r}^{\dagger}$ where $|r| \le N$ and $N \to \infty$. One can think of the vacuum as consisting of the

quantum state in which the band of negative energy states between $-m$ and $-E_{p_N}$ are occupied, where $N \to \infty$. All positive energy states are unoccupied and all states with energy less than $-E_{p_N}$ are unoccupied. The effect of the operator $\hat{a}_{-1,p_{r+s}}^{\dagger}\,\hat{a}_{-1,p_r}$, where $r + s > N$ and $r < N$, on $\left| \Psi_{\text{vac}}^{2N+1} \right\rangle$ is to destroy an electron in the occupied negative energy band and to create an electron in one of the unoccupied negative energy states underneath the occupied band. This new state has less energy than the initial vacuum state. Therefore hole theory includes eigenstates with less energy than the vacuum state.

## XII.  Redefining the Vacuum state

As can be seen from the prior discussion hole theory allows for existence of quantum states with less energy than the vacuum and field theory does not. Therefore field theory is not equivalent to hole theory. The difficulty with this result is that hole theory included all the conservation laws and symmetries associated with the single particle Dirac equation. In particular these include gauge invariance and the continuity equation. Therefore if hole theory and field theory are not equivalent we can no longer assume that field theory is gauge invariant or obeys the continuity equation. And, as has been shown this is indeed the case. It has been shown that the fact that there are no states with less free field energy than the vacuum state means that field theory is not gauge invariant and does not obey the continuity equation. This result is born out when calculations are made using perturbation theory. The offending non-gauge invariant terms that appear in these results must be removed to make the result physically correct. It should be stressed that these physically incorrect solutions are *mathematically* correct. Since the underlying theory is not gauge invariant the results of mathematical calculations should produce non-gauge invariant results [12].

What will be shown next is that it is possible to define the vacuum in field theory so that the equivalence between hole theory and field theory is restored. Refer to the definition of the vacuum state vector $\left| 0 \right\rangle$ given by (5.1). According to this definition $\left| 0 \right\rangle$ is the state in which all negative energy states are occupied by a single electron and all positive energy states are unoccupied. The top of the negative energy band has an energy of $-m$. Therefore for $|0\rangle$ all energy states with energy less than $-m$ are occupied.

Define the state vector $|0_c\rangle$ as the quantum state in which all positive energy states are unoccupied, each negative energy state in the band of states between $-m$ and $-E_c$ is occupied, and negative energy states with energy less than $-E_c$ are unoccupied where $E_c \rightarrow \infty$. The state $|0_c\rangle$ is, then, defined by,

$$|0_c\rangle = \prod_{n \in \text{band}} \hat{a}_n^\dagger |0, \text{bare}\rangle \qquad (12.1)$$

where the notation $n \in \text{band}$ means the product is taken over the band of states whose energy is between $-m$ and $-E_c$ where $E_c \rightarrow \infty$. This can also be expressed as,

$$\hat{a}_n |0_c\rangle = 0 \text{ for } \lambda_n E_{\vec{p}_n} > m$$
$$\hat{a}_n^\dagger |0_c\rangle = 0 \text{ for } -m \geq \lambda_n E_{\vec{p}_n} \geq -E_c \qquad (12.2)$$
$$\hat{a}_n |0_c\rangle = 0 \text{ for } -E_c > \lambda_n E_{\vec{p}_n}$$

Now let us take $|0_c\rangle$ as the vacuum state instead of $|0\rangle$. Compare the definition of $|0_c\rangle$ (equation (12.1)) with $|0\rangle$ (equation (5.1)). Note the state $|0_c\rangle$ is almost identical to $|0\rangle$ with the exception that for $|0_c\rangle$ the bottom of the negative energy band is handled by a limiting process. The cutoff energy $-E_c$ is assumed to finite and is taken to negative infinity at the end of a calculation. States with less energy then $-E_c$ are unoccupied. If one of these states becomes occupied then the new state will have less free field energy then $|0_c\rangle$. Therefore $|0_c\rangle$ is no longer the lower bound to the free field energy. In fact there is no lower bound to the free field energy. Therefore using $\left|0_c\right\rangle$ as the vacuum state meets the requirement of Section V where it was shown that, in order for field theory to be gauge invariant and obey the continuity equation, there must be no lower bound to the free field energy. Also, as we will show below, if $\left|0_c\right\rangle$ is used as the vacuum state then the Schwinger term is zero.

Recall the discussion in Section VIII where we have shown that quantity $I\left(\vec{y},\vec{x}\right)=\left\langle 0\left|ST\left(\vec{y},\vec{x}\right)\right|0\right\rangle$ was non-zero which meant that the Schwinger term $ST\left(\vec{y},\vec{x}\right)$ was non-zero. Here we evaluate the quantity defined by

$$I_c\left(\vec{y},\vec{x}\right)=\left\langle 0_c\left|ST\left(\vec{y},\vec{x}\right)\right|0_c\right\rangle=\left\langle 0_c\left|\left[\hat{\rho}\left(\vec{y}\right),\hat{\vec{J}}\left(\vec{x}\right)\right]\right|0_c\right\rangle \tag{12.3}$$

Use (4.17) and (4.18) in the above to yield,

$$I_c\left(\vec{y},\vec{x}\right)=q^2\sum_{nmrs}\left\langle 0_c\left|\left[\hat{a}_n^\dagger\hat{a}_m,\hat{a}_r^\dagger\hat{a}_s\right]\right|0_c\right\rangle\left(\varphi_n^{(0)\dagger}\left(\vec{y}\right)\varphi_m^{(0)}\left(\vec{y}\right)\right)\left(\varphi_r^{(0)\dagger}\left(\vec{x}\right)\vec{\alpha}\varphi_s^{(0)}\left(\vec{x}\right)\right) \tag{12.4}$$

Use (4.9) in the above to obtain,

$$I_c\left(\vec{y},\vec{x}\right)=q^2\sum_{nmrs}\left\langle 0_c\left|\begin{pmatrix}\delta_{mr}\hat{a}_n^\dagger\hat{a}_s\\-\delta_{ns}\hat{a}_r^\dagger\hat{a}_m\end{pmatrix}\right|0_c\right\rangle\left(\varphi_n^{(0)\dagger}\left(\vec{y}\right)\varphi_m^{(0)}\left(\vec{y}\right)\right)\left(\varphi_r^{(0)\dagger}\left(\vec{x}\right)\vec{\alpha}\varphi_s^{(0)}\left(\vec{x}\right)\right) \tag{12.5}$$

Use (12.2) in the above and redefine some of the dummy variables to yield,

$$I_c\left(\vec{y},\vec{x}\right)=q^2\sum_{s\in band}\sum_r\left\{\begin{matrix}\left(\varphi_s^{(0)\dagger}\left(\vec{y}\right)\varphi_r^{(0)}\left(\vec{y}\right)\right)\left(\varphi_r^{(0)\dagger}\left(\vec{x}\right)\vec{\alpha}\varphi_s^{(0)}\left(\vec{x}\right)\right)\\-\left(\varphi_s^{(0)\dagger}\left(\vec{x}\right)\vec{\alpha}\varphi_r^{(0)}\left(\vec{x}\right)\right)\left(\varphi_r^{(0)\dagger}\left(\vec{y}\right)\varphi_s^{(0)}\left(\vec{y}\right)\right)\end{matrix}\right\} \tag{12.6}$$

The notation $s\in band$ means the index 's' is summed over the states whose energy is in the band from $-m$ to $-E_c$. Note that the summation over 'r' is over all states.

Take the summation over 'r' and use (2.16) in the above to obtain,

$$I_c\left(\vec{y},\vec{x}\right) = q^2 \sum_{s\in band}\left\{\begin{array}{l}\left(\varphi_s^{(0)\dagger}\left(\vec{y}\right)\vec{\alpha}\varphi_s^{(0)}\left(\vec{x}\right)\right)\delta^{(3)}\left(\vec{x}-\vec{y}\right)\\-\left(\varphi_s^{(0)\dagger}\left(\vec{x}\right)\vec{\alpha}\varphi_s^{(0)}\left(\vec{y}\right)\right)\delta^{(3)}\left(\vec{x}-\vec{y}\right)\end{array}\right\} \qquad (12.7)$$

Next use the relationship,

$$f\left(\vec{y}\right)\delta^{(3)}\left(\vec{x}-\vec{y}\right) = f\left(\vec{x}\right) \qquad (12.8)$$

to obtain,

$$I_c\left(\vec{y},\vec{x}\right) = q^2 \sum_{s\in band}\delta^{(3)}\left(\vec{x}-\vec{y}\right)\left\{\begin{array}{l}\left(\varphi_s^{(0)\dagger}\left(\vec{x}\right)\vec{\alpha}\varphi_s^{(0)}\left(\vec{x}\right)\right)\\-\left(\varphi_s^{(0)\dagger}\left(\vec{x}\right)\vec{\alpha}\varphi_s^{(0)}\left(\vec{x}\right)\right)\end{array}\right\} = 0 \qquad (12.9)$$

Therefore the quantity $\left\langle 0_c\left|ST\left(\vec{y},\vec{x}\right)\right|0_c\right\rangle$ is zero. This allows for the possibility that the Schwinger term is zero. Now the Schwinger term is zero if all quantities of the form $\left\langle k'\left|ST\left(\vec{y},\vec{x}\right)\right|k\right\rangle$ are zero where $|k\rangle$ and $|k'\rangle$ are energy eigenstates. To show that this is the case evaluate,

$$\left\langle k'\left|ST\left(\vec{y},\vec{x}\right)\right|k\right\rangle = q^2\sum_{nmrs}\left\langle k'\left|\left[\hat{a}_n^\dagger\hat{a}_m,\hat{a}_r^\dagger\hat{a}_s\right]\right|k\right\rangle\left(\varphi_n^{(0)\dagger}\left(\vec{y}\right)\varphi_m^{(0)}\left(\vec{y}\right)\right)\left(\varphi_r^{(0)\dagger}\left(\vec{x}\right)\vec{\alpha}\varphi_s^{(0)}\left(\vec{x}\right)\right) \qquad (12.10)$$

Next use (4.9) in the above to yield,

$$\left\langle k'\left|ST\left(\vec{y},\vec{x}\right)\right|k\right\rangle = q^2\sum_{nmrs}\left\langle k'\left|\left(\begin{array}{l}\delta_{mr}\hat{a}_n^\dagger\hat{a}_s\\-\delta_{ns}\hat{a}_r^\dagger\hat{a}_m\end{array}\right)\right|k\right\rangle\left(\varphi_n^{(0)\dagger}\left(\vec{y}\right)\varphi_m^{(0)}\left(\vec{y}\right)\right)\left(\varphi_r^{(0)\dagger}\left(\vec{x}\right)\vec{\alpha}\varphi_s^{(0)}\left(\vec{x}\right)\right) \qquad (12.11)$$

This becomes,

$$\langle k'|ST(\vec{y},\vec{x})|k\rangle = q^2 \left( \begin{array}{l} \sum\limits_{nms} \langle k'|\left(\hat{a}_n^\dagger \hat{a}_s\right)|k\rangle \left(\varphi_n^{(0)\dagger}(\vec{y})\varphi_m^{(0)}(\vec{y})\right)\left(\varphi_m^{(0)\dagger}(\vec{x})\vec{\alpha}\varphi_s^{(0)}(\vec{x})\right) \\ -\sum\limits_{nmr} \langle k'|\left(\hat{a}_r^\dagger \hat{a}_m\right)|k\rangle \left(\varphi_r^{(0)\dagger}(\vec{x})\vec{\alpha}\varphi_n^{(0)}(\vec{x})\right)\left(\varphi_n^{(0)\dagger}(\vec{y})\varphi_m^{(0)}(\vec{y})\right) \end{array} \right)$$

$$(12.12)$$

Next use (2.16) and (12.8) redefine some dummy variables to obtain,

$$\langle k'|ST(\vec{y},\vec{x})|k\rangle = q^2 \delta(\vec{x}-\vec{y}) \left( \begin{array}{l} \sum\limits_{ns} \langle k'|\left(\hat{a}_n^\dagger \hat{a}_s\right)|k\rangle \left(\varphi_n^{(0)\dagger}(\vec{y})\vec{\alpha}\varphi_s^{(0)}(\vec{x})\right) \\ -\sum\limits_{ns} \langle k'|\left(\hat{a}_n^\dagger \hat{a}_s\right)|k\rangle \left(\varphi_n^{(0)\dagger}(\vec{x})\vec{\alpha}\varphi_s^{(0)}(\vec{y})\right) \end{array} \right)$$

$$= q^2 \delta(\vec{x}-\vec{y}) \sum\limits_{ns} \left( \begin{array}{l} \langle k'|\left(\hat{a}_n^\dagger \hat{a}_s\right)|k\rangle \left(\varphi_n^{(0)\dagger}(\vec{x})\vec{\alpha}\varphi_s^{(0)}(\vec{x})\right) \\ -\langle k'|\left(\hat{a}_n^\dagger \hat{a}_s\right)|k\rangle \left(\varphi_n^{(0)\dagger}(\vec{x})\vec{\alpha}\varphi_s^{(0)}(\vec{x})\right) \end{array} \right) = 0$$

$$(12.13)$$

Therefore if $|0_c\rangle$ is used as the vacuum state then the Schwinger term

$ST(\vec{y},\vec{x}) = 0$ because all quantities $\langle k'|ST(\vec{y},\vec{x})|k\rangle$ are zero. And, as has been shown,

this is a necessary condition for quantum field theory to be gauge invariant and for the continuity equation to be true.

## XIII. Conclusion

It has been shown that hole theory and the Schrödinger representation of quantum

field theory are not equivalent. This is due to the fact that, in field theory, there can be no

states with less free field energy then the vacuum state $|0\rangle$, however, as was shown in

Section X, this is not the case for hole theory. In hole theory it is possible to extract energy from the vacuum state through interaction with an appropriately applied electric field.

As a result of this lack of equivalence between the two theories we must examine

whether the conservation laws and symmetries that are associated with the single particle

Dirac equation are valid in field theory. It is seen that they are not. Field theory is not gauge invariant and the continuity equation does not hold. This is due to the fact that the Schwinger term is not zero. And the Schwinger term is not zero because field theory does not allow for the existence of states with less free field energy than the vacuum state $|0\rangle$.

In order to restore the equivalence between hole theory and field theory it is necessary to modify the definition of the vacuum state. Instead of the using the state vector $|0\rangle$ for the vacuum the state vector $|0_c\rangle$ is used instead. When $|0_c\rangle$ is used there can exist quantum states with less free field energy than the vacuum which is a necessary requirement for the Schwinger term to be zero. It is shown that the Schwinger term is zero when $|0_c\rangle$ is used as the vacuum state. In this case field theory will be gauge invariance and the continuity equation will be valid.

## **Appendix A**

In order to use perturbation theory we must convert from the Schrödinger picture to the interaction picture (see Chapter 4-2 of [17]). Refer back to equation (4.20). Write the Hamiltonian as,

$$\hat{H} = \hat{H}_0 + \hat{V} \tag{A.1}$$

where $\hat{H}_0$ is the unperturbed free field Hamiltonian and $\hat{V}$ is the perturbation. From (4.22),

$$\hat{V} = -\int \hat{J}(\vec{x}) \cdot \vec{A}(\vec{x}, t) d\vec{x} + \int \rho(\vec{x}) \cdot A_0(\vec{x}, t) d\vec{x} \tag{A.2}$$

Define the interaction state vector by,

$$|\Omega_I\rangle = e^{i\hat{H}_0 t} |\Omega\rangle \tag{A.3}$$

Interaction operators $\hat{O}_I$ are defined in terms of Schrödinger operators $\hat{O}$ according to,

$$\hat{O}_I = e^{i\hat{H}_0 t} \hat{O} e^{-i\hat{H}_0 t} \tag{A.4}$$

The expectation values of operators have the same value in both representations, i.e.,

$O_e = \left\langle \Omega_I \middle| \hat{O}_I \middle| \Omega_I \right\rangle = \left\langle \Omega \middle| \hat{O} \middle| \Omega \right\rangle$. Use the above expressions in (4.20) to obtain,

$$i \frac{\partial \left| \Omega_I \right\rangle}{\partial t} = \hat{V}_I \left| \Omega_I \right\rangle \tag{A.5}$$

A formal solution of above is given by,

$$\left| \Omega_I(t) \right\rangle = \left( 1 - i \int_{t_0}^{t} \hat{V}_I(t_1) dt_1 + (-i)^2 \int_{t_0}^{t} \hat{V}_I(t_1) dt_1 \int_{t_0}^{t_1} \hat{V}_I(t_2) dt_2 \right) \left| \Omega_I(t_0) \right\rangle \tag{A.6}$$

The current expectation value is then given by,

$$\vec{J}_e(\vec{x}, t) = \left\langle \Omega_I(t) \middle| \hat{\vec{J}}_I(\vec{x}, t) \middle| \Omega_I(t) \right\rangle$$

$$= \left\langle \Omega_I(t_0) \middle| \hat{\vec{J}}_I(\vec{x}, t) \middle| \Omega_I(t_0) \right\rangle - i \left\langle \Omega_I(t_0) \middle| \left[ \hat{\vec{J}}_I(\vec{x}, t), \int_{t_0}^{t} \hat{V}_I(t_1) dt_1 \right] \middle| \Omega_I(t_0) \right\rangle + O\left(V^2\right)$$

$$\tag{A.7}$$

Let the initial state, at time $t = t_0$, be the vacuum state $\left| 0 \right\rangle$. The first order vacuum current at time t is then,

$$\vec{J}_{vac}(\vec{x}, t) \simeq \left\langle 0_I \middle| \hat{\vec{J}}_I(\vec{x}, t) \middle| 0_I \right\rangle - i \left\langle 0_I \middle| \left[ \hat{\vec{J}}_I(\vec{x}, t), \int_{t_0}^{t} \hat{V}_I(t_1) dt_1 \right] \middle| 0_I \right\rangle \tag{A.8}$$

where $\left| 0_I \right\rangle$ is the interaction vacuum state and is given by,

$$\left| 0_I \right\rangle = e^{i\hat{H}_0 t} \left| 0 \right\rangle = \left| 0 \right\rangle \tag{A.9}$$

It is easy to show that $\left\langle 0_I \middle| \hat{\vec{J}}_I(\vec{x}, t) \middle| 0_I \right\rangle = 0$. Use this, along with (A.9) and let the initial time $t_0 = -\infty$ to obtain equation (9.8) in the text.

## **Appendix B**

Expand $\varphi_n\left(\vec{x}, t\right)$ as a series in the charge 'q',

$$\varphi_n\left(\vec{x}, t\right) = \varphi_n^{(0)}\left(\vec{x}, t\right) + q\varphi_n^{(1)}\left(\vec{x}, t\right) + q^2\varphi_n^{(2)}\left(\vec{x}, t\right) + O\left(q^3\right) \tag{B.1}$$

where $O\left(q^3\right)$ means terms to the third order in the perturbation or higher. The free field energy of the state $\varphi_n\left(\vec{x}, t\right)$ is,

$$\xi_f\left(\varphi_n\right) \equiv \left\langle \varphi_n^\dagger H_0 \varphi_n \right\rangle = \left\langle \left(\varphi_n^{(0)\dagger} + q\varphi_n^{(1)\dagger} + q^2\varphi_n^{(2)\dagger}\right) H_0 \left(\varphi_n^{(0)} + q\varphi_n^{(1)} + q^2\varphi_n^{(2)}\right) \right\rangle + O\left(q^3\right)$$

$$\tag{B.2}$$

Rearrange terms to obtain,

$$\begin{aligned}\xi_f\left(\varphi_n\right) = \left\langle \varphi_n^{(0)\dagger} H_0 \varphi_n^{(0)} \right\rangle &+ q\left(\left\langle \varphi_n^{(0)\dagger} H_0 \varphi_n^{(1)} \right\rangle + \left\langle \varphi_n^{(1)\dagger} H_0 \varphi_n^{(0)} \right\rangle\right) \\ &+ q^2\left(\left\langle \varphi_n^{(1)\dagger} H_0 \varphi_n^{(1)} \right\rangle + \left\langle \varphi_n^{(0)\dagger} H_0 \varphi_n^{(2)} \right\rangle + \left\langle \varphi_n^{(2)\dagger} H_0 \varphi_n^{(0)} \right\rangle\right) + O\left(q^3\right)\end{aligned} \tag{B.3}$$

Use $H_0 \varphi_n^{(0)} = \lambda_n E_n \varphi_n^{(0)}$ in the above to obtain,

$$\xi_f\left(\varphi_n\right) = q^2 \left\langle \varphi_n^{(1)\dagger} H_0 \varphi_n^{(1)} \right\rangle + \lambda_n E_n \left( \begin{matrix} \left\langle \varphi_n^{(0)\dagger} \varphi_n^{(0)} \right\rangle + q\left(\left\langle \varphi_n^{(0)\dagger} \varphi_n^{(1)} \right\rangle + \left\langle \varphi_n^{(1)\dagger} \varphi_n^{(0)} \right\rangle\right) \\ + q^2\left(\left\langle \varphi_n^{(0)\dagger} \varphi_n^{(2)} \right\rangle + \left\langle \varphi_n^{(2)\dagger} \varphi_n^{(0)} \right\rangle\right) \end{matrix} \right) + O\left(q^3\right)$$

$$\tag{B.4}$$

The Dirac equation does not affect the normalization condition therefore,

$$\left\langle \varphi_n^\dagger \varphi_n \right\rangle = 1 \tag{B.5}$$

Use (B.1) in the above to obtain,

$$1 = \left\langle \varphi_n^{(0)\dagger} \varphi_n^{(0)} \right\rangle + q \left( \begin{matrix} \left\langle \varphi_n^{(1)\dagger} \varphi_n^{(0)} \right\rangle \\ + \left\langle \varphi_n^{(0)\dagger} \varphi_n^{(1)} \right\rangle \end{matrix} \right) + q^2 \left( \begin{matrix} \left\langle \varphi_n^{(1)\dagger} \varphi_n^{(1)} \right\rangle + \left\langle \varphi_n^{(0)\dagger} \varphi_n^{(2)} \right\rangle \\ + \left\langle \varphi_n^{(2)\dagger} \varphi_n^{(0)} \right\rangle \end{matrix} \right) + O\left(q^3\right) \tag{B.6}$$

Rearrange terms to yield,

$$1 - q^2 \left\langle \varphi_n^{(1)\dagger} \varphi_n^{(1)} \right\rangle = \left\langle \varphi_n^{(0)\dagger} \varphi_n^{(0)} \right\rangle + q \begin{pmatrix} \left\langle \varphi_n^{(1)\dagger} \varphi_n^{(0)} \right\rangle \\ + \left\langle \varphi_n^{(0)\dagger} \varphi_n^{(1)} \right\rangle \end{pmatrix} + q^2 \begin{pmatrix} \left\langle \varphi_n^{(0)\dagger} \varphi_n^{(2)} \right\rangle \\ + \left\langle \varphi_n^{(2)\dagger} \varphi_n^{(0)} \right\rangle \end{pmatrix} + O\left(q^3\right) \quad \text{(B.7)}$$

Use this in (B.4) to obtain,

$$\xi_f\left(\varphi_n\right) = q^2 \left\langle \varphi_n^{(1)\dagger} H_0 \varphi_n^{(1)} \right\rangle + \lambda_n E_n \left(1 - q^2 \left\langle \varphi_n^{(1)\dagger} \varphi_n^{(1)} \right\rangle \right) + O\left(q^3\right) \tag{B.8}$$

Rearrange terms to obtain,

$$\xi_f\left(\varphi_n\right) = \lambda_n E_n + q^2 \left( \left\langle \varphi_n^{(1)\dagger} H_0 \varphi_n^{(1)} \right\rangle - \lambda_n E_n \left\langle \varphi_n^{(1)\dagger} \varphi_n^{(1)} \right\rangle \right) + O\left(q^3\right) \tag{B.9}$$

Therefore the change in the free field energy of state $\varphi_n$ is,

$$\Delta \xi_{f,n}\left(t_i \to t_f\right) = \xi_f\left(\varphi_n\right) - \lambda_n E_n = q^2 \left( \left\langle \varphi_n^{(1)\dagger} H_0 \varphi_n^{(1)} \right\rangle - \lambda_n E_n \left\langle \varphi_n^{(1)\dagger} \varphi_n^{(1)} \right\rangle \right) + O\left(q^3\right)$$
$$\tag{B.10}$$

which is (10.7) in the text.

To evaluate the above we have to obtain $\varphi_n^{(1)}$. Based on the discussion in Appendix A this is given by,

$$\varphi_n^{(1)}\left(\vec{x}, t\right) = -iq e^{-iH_0 t} \int_{t_i}^{t} dt' V_I\left(\vec{x}, t'\right) \varphi_n^{(0)}\left(\vec{x}, 0\right) \tag{B.11}$$

In the above expression $V_I$ is given by $V_I = e^{-iH_0 t} V e^{+iH_0 t}$ where $V = \left(-\vec{\alpha} \cdot \vec{A} + A_0\right)$. Use this in the above to obtain,

$$\varphi_n^{(1)}\left(\vec{x}, t\right) = -i \int_{-\infty}^{t} dt' e^{-iH_0(t-t')} V\left(\vec{x}, t'\right) \varphi_n^{(0)}\left(\vec{x}, t'\right) \tag{B.12}$$

Next expand the quantity $V\left(\vec{x}, t'\right) \varphi_n^{(0)}\left(\vec{x}, t'\right)$ in terms of the basis states $\varphi_s^{(0)}\left(\vec{x}, t'\right)$ to obtain,

$$e^{iH_0(t-t')}V(\vec{x},t')\varphi_n^{(0)}(\vec{x},t')$$
$$= \sum_s \left(e^{-iH_0(t-t')}\varphi_s^{(0)}(\vec{x},t')\right)\int \varphi_s^{(0)\dagger}(\vec{x}',t')\left(V(\vec{x}',t')\varphi_n^{(0)}(\vec{x}',t')\right)d\vec{x}' \quad \text{(B.13)}$$

Use this and $e^{-iH_0(t-t')}\varphi_s^{(0)}(\vec{x},t') = \varphi_s^{(0)}(\vec{x},t)$ to obtain,

$$\varphi_n^{(1)}(\vec{x},t) = -i\int_{-\infty}^{t}dt'\sum_s\varphi_s^{(0)}(\vec{x},t)\int \varphi_s^{(0)\dagger}(\vec{x}',t')V(\vec{x}',t')\varphi_n^{(0)}(\vec{x}',t')d\vec{x}' \quad \text{(B.14)}$$

## **Appendix C**

From (B.14) and the discussion of the 1-1D Dirac equation we obtain,

$$\varphi_{\lambda,p_r}^{(1)}(z,t=+\infty) = -i\sum_{\lambda'=\pm1}\sum_{p_s}\varphi_{\lambda',p_s}^{(0)}(z,t)\int_{-\infty}^{+\infty}dt'\int dz'\varphi_{\lambda',p_s}^{(0)\dagger}(z',t')A_0(z',t')\varphi_{\lambda,p_r}^{(0)}(z',t')$$

$$\text{(C.1)}$$

for the case where $A_z = 0$. Next use (10.11) in the above to obtain,

$$\varphi_{\lambda,p_r}^{(1)}(z,t) = -i\sum_{\lambda'=\pm1}\sum_{p_s}\varphi_{\lambda',p_s}^{(0)}(z,t)\int_{-\infty}^{t}dt'\int dz'A_0(z',t')u_{\lambda',p_s}^{\dagger}u_{\lambda,p_r}e^{-i\left(\lambda E_{p_r}-\lambda'E_{p_s}\right)t'}e^{i(p_r-p_s)z'}$$

$$\text{(C.2)}$$

Use (10.14) in the above to yield,

$$\varphi_{\lambda,p_r}^{(1)}(z,t=\infty) = -i\sum_{\lambda'=\pm1}\sum_{p_s}\varphi_{\lambda',p_s}^{(0)}(z,t)u_{\lambda',p_s}^{\dagger}u_{\lambda,p_r}\int_{-\infty}^{+\infty}dt'\int dz'\left(\begin{array}{c}e^{-i\left(\lambda E_{p_r}-\lambda'E_{p_s}\right)t'}e^{i(p_r-p_s)z'}\\ \times\left(e^{ikz'}+e^{-ikz'}\right)\int_{-m}^{+m}e^{iqt'}dq\end{array}\right)$$

$$\text{(C.3)}$$

Integrate over $z'$ and $t'$ to obtain,

$$\varphi_{\lambda,p_r}^{(1)}(z,t=\infty) = -4\pi^2i\sum_{\lambda'=\pm1}\sum_{p_s}\left\{\begin{array}{c}\varphi_{\lambda',p_s}^{(0)}(z,t)u_{\lambda',p_s}^{\dagger}u_{\lambda,p_r}\\ \times\left(\begin{array}{c}\delta\left(k+p_r-p_s\right)\\ +\delta\left(-k+p_r-p_s\right)\end{array}\right)\int_{-m}^{+m}\delta\left(\begin{array}{c}-\lambda E_{p_r}\\ +\lambda'E_{p_s}+q\end{array}\right)dq\end{array}\right\} \quad \text{(C.4)}$$

Make the substitution,

$$\sum_{p_s} \rightarrow \int\limits_{-\infty}^{+\infty} \frac{L dp_s}{2\pi} \tag{C.5}$$

in the above to obtain,

$$\varphi_{\lambda,p_r}^{(1)}\left(z,t=\infty\right) = -2\pi Li \sum_{\lambda'=\pm 1} \int dp_s \left\{ \begin{array}{l} \varphi_{\lambda',p_s}^{(0)}\left(z,t\right) u_{\lambda',p_s}^{\dagger} u_{\lambda,p_r} \\ \times \left(\begin{array}{l} \delta\left(k+p_r-p_s\right) \\ +\delta\left(-k+p_r-p_s\right) \end{array}\right)_{-m}^{+m} \delta\left(\begin{array}{l} q-\lambda E_{p_r} \\ +\lambda' E_{p_s} \end{array}\right) dq \end{array} \right\} \tag{C.6}$$

Next integrate with respect to momentum $p_s$ to obtain,

$$\varphi_{\lambda,p_r}^{(1)}\left(z,t=\infty\right) = -2\pi Li \sum_{\lambda'=\pm 1} \left\{ \begin{array}{l} \varphi_{\lambda',p_r+k}^{(0)}\left(z,t\right) u_{\lambda',p_r+k}^{\dagger} u_{\lambda,p_r} \int\limits_{-m}^{+m} \delta\left(\begin{array}{l} q-\lambda E_{p_r} \\ +\lambda' E_{p_r+k} \end{array}\right) dq \\ +\left(k \rightarrow -k\right) \end{array} \right\} \tag{C.7}$$

Now for sufficiently small k ( $k < m$ ) we have that,

$$\int\limits_{-m}^{+m} \delta\left(q-\lambda E_{p_r}+\lambda' E_{p_r+k}\right) dq = \delta\left(\lambda-\lambda'\right) \tag{C.8}$$

Use this in (C.7) to obtain,

$$\varphi_{\lambda,p_r}^{(1)}\left(z,t=\infty\right) = -2\pi Li \left\{ \left(\varphi_{\lambda,p_r+k}^{(0)}\left(z,t\right) u_{\lambda,p_r+k}^{\dagger} u_{\lambda,p_r}\right) + \left(k \rightarrow -k\right) \right\} \tag{C.9}$$

Define,

$$c_{\lambda,p_r,k} = 2\pi L u_{\lambda,p_r+k}^{\dagger} u_{\lambda,p_r} \tag{C.10}$$

Use this in (C.9) to obtain,

$$\varphi_{\lambda,p_r}^{(1)} = -i \left\{ c_{\lambda,p_r,k} \varphi_{\lambda,p_r+k}^{(0)} + c_{\lambda,p_r,-k} \varphi_{\lambda,p_r-k}^{(0)} \right\} \tag{C.11}$$

## **Appendix D**

We want to evaluate equation (10.20) and show that it results in equation (10.21). Use (10.16) and (10.12) to obtain,

$$c_{\lambda,p_r,k} = 2\pi \sqrt{\frac{\lambda E_{p_r+k} + m}{2\lambda E_{p_r+k}}} \sqrt{\frac{\lambda E_{p_r} + m}{2\lambda E_{p_r}}} \left( 1 + \frac{p_r(p_r+k)}{\left(\lambda E_{p_r+k} + m\right)\left(\lambda E_{p_r} + m\right)} \right) \qquad (D.1)$$

Use $\left(\lambda E_p\right)^2 - m^2 = p^2$ in the above to obtain,

$$c_{\lambda,p_r,k} = 2\pi \sqrt{\frac{\lambda E_{p_r+k} + m}{2\lambda E_{p_r+k}}} \sqrt{\frac{\lambda E_{p_r} + m}{2\lambda E_{p_r}}} \left( 1 + \frac{\left(\lambda E_{p_r+k} - m\right)\left(\lambda E_{p_r} - m\right)}{p_r(p_r+k)} \right) \qquad (D.2)$$

Use this result to yield,

$$\left| c_{\lambda,p_r,k} \right|^2 = 4\pi^2 \left( \frac{\left(\lambda E_{p_r+k}\right)\left(\lambda E_{p_r}\right) + p_r(p_r+k) + m^2}{2\left(\lambda E_{p_r+k}\right)\left(\lambda E_{p_r}\right)} \right) \qquad (D.3)$$

Use this in (10.20) along with the fact that $\lambda^2 = 1$ to obtain

$$\Delta\xi_f\left(\lambda, p_r; t_i \rightarrow t_f\right) = 2\pi^2 q^2 \lambda \left( \begin{array}{l} \left(1 + \dfrac{p_r(p_r+k) + m^2}{E_{p_r+k} E_{p_r}}\right)\left(E_{p_r+k} - E_{p_r}\right) \\ + \left(1 + \dfrac{p_r(p_r-k) + m^2}{E_{p_r-k} E_{p_r}}\right)\left(E_{p_r-k} - E_{p_r}\right) \end{array} \right) + O\left(q^3\right)$$

$$(D.4)$$

This yields,

$$\Delta\xi_f\left(\lambda, p_r; t_i \rightarrow t_f\right) = 2\pi^2 q^2 \lambda \left( \begin{array}{l} \left(1 + \dfrac{E_{p_r+k}}{E_{p_r}} - \dfrac{k(p_r+k)}{E_{p_r+k} E_{p_r}}\right)\left(E_{p_r+k} - E_{p_r}\right) \\ + \left(k \rightarrow -k\right) \end{array} \right) + O\left(q^3\right)$$

$$(D.5)$$

Some additional algebraic manipulation yields,

$$\Delta \xi_f \left( \lambda, p_r; t_i \to t_f \right) = 2\pi^2 q^2 \lambda \left( \begin{array}{l} \left( -E_{p_r} + \dfrac{E_{p_r+k}^2}{E_{p_r}} - \dfrac{k\left(p_r+k\right)}{E_{p_r}} + \dfrac{k\left(p_r+k\right)}{E_{p_r+k}} \right) \\ + \left( k \to -k \right) \end{array} \right) + O\left( q^3 \right)$$

(D.6)

Use some simple algebra to obtain,

$$\Delta \xi_f \left( \lambda, p_r; t_i \to t_f \right) = 2\pi^2 q^2 \lambda \left( \left( \dfrac{p_r k}{E_{p_r}} + \dfrac{k\left(p_r+k\right)}{E_{p_r+k}} \right) + \left( \dfrac{-p_r k}{E_{p_r}} - \dfrac{k\left(p_r-k\right)}{E_{p_r-k}} \right) \right) + O\left( q^3 \right)$$

(D.7)

Use this result to yield (10.21).

## Appendix E

Assume k is positive. Then it can be shown that,

$$\frac{\left( p_r + k \right)}{E_{p_r+k}} > \frac{\left( p_r - k \right)}{E_{p_r-k}} \text{ for all } p_r$$

(E.1)

First consider the case where $p_r$ is positive. The relationship is obviously true for $k > p_r$. Now let $p_r > k$. In this case both sides of (E.1) are positive therefore we can square both sides to obtain,

$$\left( p_r + k \right)^2 E_{p_r-k}^2 > \left( p_r - k \right)^2 E_{p_r+k}^2$$

(E.2)

From this we obtain,

$$\left( p_r + k \right)^2 \left( \left( p_r - k \right)^2 + m^2 \right) > \left( p_r - k \right)^2 \left( \left( p_r + k \right)^2 + m^2 \right)$$

(E.3)

This yields,

$$\left( p_r + k \right)^2 > \left( p_r - k \right)^2$$

(E.4)

which is true for positive k and $p_r > k$. If $p_r$ is negative then (E.1) becomes,

$$\frac{\left(-|p_r|+k\right)}{E_{|p_r|-k}} > \frac{\left(-|p_r|-k\right)}{E_{|p_r|+k}} \tag{E.5}$$

This yields,

$$\frac{\left(|p_r|+k\right)}{E_{|p_r|+k}} > \frac{\left(|p_r|-k\right)}{E_{|p_r|-k}} \tag{E.6}$$

which is obviously true from the previous discussion.